\documentclass[preprint,showpacs,superscriptaddress,nofootinbib]{revtex4}

\usepackage{amsmath,amssymb}

\newcommand{\TBM}{{\text{TB}}}
\newcommand{\MNS}{{\text{MNS}}}

\newcommand{\tr}{{\text{Tr}}}
\newcommand{\dprime}{{\prime\prime}}
\newcommand{\eV}{{\text{eV}}}
\newcommand{\MeV}{{\text{MeV}}}
\newcommand{\GeV}{{\text{GeV}}}
\newcommand{\BR}{\text{BR}}
\newcommand{\meee}{\mu \to \bar{e}ee}
\newcommand{\meg}{\mu \to e \gamma}
\newcommand{\llg}{\ell \to \ell^\prime \gamma}
\newcommand{\tlll}{\tau \to \overline{\ell}\ell^\prime \ell^\dprime}

\newcommand{\temm}{\tau \to \overline{e}\mu\mu}
\newcommand{\tLeLmLmL}{\tau_L \to \overline{e_L}\mu_L\mu_L}
\newcommand{\tmee}{\tau \to \overline{\mu}ee}
\newcommand{\tLmLeLeL}{\tau_L \to \overline{\mu_L}e_Le_L}

\begin{document}
\title{
 Phenomenology in the Higgs Triplet Model with the $A_4$ Symmetry
}

%%%%%%%%%%%%%%%%%%%%%%%%%%%%%%%%%%%%%%%%%%%%%%%%%%%%%%%%%%%%%%%%%%%%%%
\author{Takeshi Fukuyama}
\email{fukuyama@se.ritsumei.ac.jp}
\affiliation{Department of Physics and R-GIRO,
Ritsumeikan University, Kusatsu, Shiga,
525-8577, Japan}
\affiliation{Maskawa Institute for Science and Culture,
Kyoto Sangyo University, Kyoto 603-8555, Japan}
%%%%%%%%%%%%%%%%%%%%%%%%%%%%%%%%%%%%%%%%%%%%%%%%%%%%%%%%%%%%%%%%%%%%%%
\author{Hiroaki Sugiyama}
\email{hiroaki@fc.ritsumei.ac.jp}
\affiliation{Department of Physics and R-GIRO,
Ritsumeikan University, Kusatsu, Shiga,
525-8577, Japan}
%%%%%%%%%%%%%%%%%%%%%%%%%%%%%%%%%%%%%%%%%%%%%%%%%%%%%%%%%%%%%%%%%%%%%%%
\author{Koji Tsumura}
\email{ktsumura@ictp.it}
\affiliation{The Abdus Salam ICTP of UNESCO and IAEA,
Strada Costiera 11, 34151 Trieste, Italy}

\begin{abstract}
 We discuss the phenomenology of
doubly and singly charged Higgs bosons
(of $SU(2)_L$-triplet fields)
in the simplest $A_4$-symmetric version
of the Higgs Triplet Model.
 Mass eigenstates of these Higgs bosons
are obtained explicitly
from the Higgs potential.
 It is shown that
their decays into a pair of leptons
have unique flavor structures
which can be tested at the LHC
if some of their masses are below the TeV scale.
 Sizable decay rates for $\tmee$ and $\temm$
can be obtained naturally
while other $\tlll$, $\meee$,
and $\ell\to \ell^\prime \gamma$
are almost forbidden in this model.
 Contributions of these Higgs bosons
to the non-standard interactions of neutrinos
are also considered.
\end{abstract}

\preprint{IC/2010/022}

\pacs{11.30.Hv, 13.35.-r, 14.60.Pq, 14.80.Cp}
%11.30.Hv 	Flavor symmetries 
%13.35.-r 	Decays of leptons
%14.60.Pq 	Neutrino mass and mixing
%14.80.Cp 	Non-standard-model Higgs bosons

\maketitle

\section{Introduction}
 Two curious features of the lepton sector
have been clarified by neutrino oscillation measurements%
~\cite{solar,atm,acc,Apollonio:2002gd,:2008ee}.
 One feature is that
neutrinos have nonzero masses
which are extremely smaller than other fermion masses.
 This seems to indicate that
neutrino masses are generated
by a different mechanism from
that for other fermions.
 In the Standard Model of particle physics~(SM),
fermion masses are obtained
with the vacuum expectation value~(vev)
of an $SU(2)_L$-doublet scalar field
while neutrinos are massless
because of the absence of the right-handed neutrinos.
 The Higgs Triplet Model~(HTM)~\cite{HTM,Schechter:1980gr}
is a simple extension of the SM
with an $SU(2)_L$-triplet Higgs boson
of hypercharge $Y=2$
whose vev provides Majorana neutrino masses
without introducing right-handed neutrinos.
 The HTM has a predictive phenomenology
because the matrix of triplet Yukawa couplings
$h_{\ell\ell^\prime}$
is proportional to the neutrino mass matrix $(M_\nu)_{\ell\ell^\prime}$
in the flavor basis
and $M_\nu$ is very restricted now by neutrino oscillation data.
 The characteristic particle in the HTM
is the doubly charged Higgs boson $H^{\pm\pm}$
which will be discovered at hadron colliders
(Tevatron and LHC)
if it is light enough.
 Tevatron has been searching for $H^{\pm\pm}$
and put lower bounds on its mass,
$m_{H^{\pm\pm}} > 112\,\text{--}\,150\,\GeV$~\cite{Tevatron},
where one of decay branching ratios (BRs)
into same-signed charged leptons
is assumed simply to be 100\%.
 If $\BR(H^{--} \to \ell\ell^\prime)$ are measured,
important information on the neutrino mass matrix
will be obtained~\cite{HTMLHC,Akeroyd:2007zv, Perez:2008ha, Nishiura}.
 Even though $H^{\pm\pm}$ is too heavy
to be produced at collider experiments,
lepton flavor violating processes
($\meee$, $\tlll$, etc.) are possible
if $h_{\ell\ell^\prime}$ are sizable.
 Previous works for dependences of lepton flavor violating processes
on the parameters in $M_\nu$
can be found in \cite{Chun:2003ej,HTMLFV}.

 The other interesting feature of the lepton sector is
the nontrivial structure of the lepton flavor mixing.
 The lepton flavors are mixed by
two large mixing angles
($\theta_{23}\simeq 45^\circ$ and
$\theta_{12}\simeq 34^\circ$)
in contrast with the structure
of the quark sector which has small mixings only.
 It seems natural to expect that
there is some underlying physics
for the special feature of the lepton flavor.
 As the candidate for that,
non-Abelian discrete symmetries
have been studied 
(See e.~g., \cite{discrete} and references therein).
 An interesting choice is the $A_4$ symmetry
because this is the minimal one
including the $3$-dimensional irreducible representation
which seems suitable for three flavors of the lepton.
 Some simple models based on the $A_4$ symmetry
can be found in e.~g.,
\cite{A4NR,Chen:2005jm,Ma:2005mw,Hirsch:2007kh,
Ma:2004zv,Hirsch:2005mc,Altarelli:2005yp,Ma:2008ym}.

 In this article,
we deal with the simplest $A_4$-symmetric version
of the Higgs Triplet Model (A4HTM).
 The mass eigenstates of doubly charged Higgs bosons
$H_i^{\pm\pm}$ are obtained explicitly
from the Higgs potential.
 We see the characteristic flavor structures
of $\BR(H_i^{--}\to \ell \ell^\prime)$.
 Other exotic processes like $\tlll$ are also considered.
 Similarly,
we investigate also phenomenology
of ``triplet-like'' singly charged Higgs bosons $H_{Ti}^\pm$;
 we refer to
the mass eigenstates
which are made mainly from triplet scalar fields
as the triple-like Higgs bosons.

 This article is organized as follows.
 Section~\ref{sect:A4HTM} is devoted to
the explanation of the A4HTM\@.
 The Higgs sector is analyzed in Sec.~\ref{sect:Higgs},
and mass eigenstates of Higgs bosons are obtained there.
 Section~\ref{sect:pheno} shows
phenomenology of the Higgs bosons:
leptonic decays of the Higgs bosons,
lepton flavor violating decays of charged leptons,
non-standard interactions of neutrinos etc.
 We consider constraints on the model in Sec.~\ref{sect:constraints}.
 Conclusions are given in Sec.~\ref{sect:concl}.
 Throughout this article,
we use the words "triplet" etc.\ only for
the representations of $SU(2)_L$
and "${\bf 3}$-representation" etc.\ for
the ones of $A_4$
in order to avoid confusion.

%%%%%%%%%%%   sect: A4HTM  %%%%%%%%%
\section{Higgs triplet model with $A_4$ symmetry}
\label{sect:A4HTM}

 The $A_4$ symmetry is characterized by
two elemental transformations $S$ and $T$
which satisfy
\begin{eqnarray}
S^2 = T^3 = (ST)^3 = 1. 
\end{eqnarray}
 There are three 1-dimensional and one 3-dimensional
irreducible representations.
 We use the following representations:
\begin{eqnarray}
{\bf 1}&:&
S\, {\bf 1} = {\bf 1}, \quad
T\, {\bf 1} = {\bf 1},\\
{\bf 1}^\prime&:&
S\, {\bf 1}^\prime = {\bf 1}^\prime, \quad
T\, {\bf 1}^\prime = \omega {\bf 1}^\prime,\\
{\bf 1}^\dprime&:&
S\, {\bf 1}^\dprime = {\bf 1}^\dprime, \quad
T\, {\bf 1}^\dprime = \omega^2 {\bf 1}^\dprime,\\
{\bf 3}&:&
S\, {\bf 3}
 = \begin{pmatrix}
    1 &  0 &  0\\
    0 & -1 &  0\\
    0 &  0 & -1
   \end{pmatrix}
   {\bf 3}, \quad
T\, {\bf 3}
 = \begin{pmatrix}
    0 & 1 & 0\\
    0 & 0 & 1\\
    1 & 0 & 0
   \end{pmatrix}
   {\bf 3} ,
\label{eq:sdiag}
\end{eqnarray}
where $\omega \equiv \text{exp}(2\pi i/3)$.
 We refer to the basis in eq.~(\ref{eq:sdiag})
as the $S$-diagonal basis.
 See appendix
for another simple choice (the ``$T$-diagonal basis'').
 Because of ${\bf 3}^\ast={\bf 3}$ in the $S$-diagonal basis,
the basis seems better than the $T$-diagonal one
for the construction of the $A_4$-symmetric Higgs potential.

\begin{table}[t]
\begin{tabular}{c||c|c|c|c|c|c|c}
 {}
 & $\psi_{1R}^-$
 & $\psi_{2R}^-$
 & $\psi_{3R}^-$
 & $\Psi_{AL}
    = \left(
       \begin{array}{c}
	\psi_{AL}^0\\
	\psi_{AL}^-
       \end{array}
      \right)$
 & $\Phi_A
    = \left(
       \begin{array}{c}
	\phi_A^+\\
	\phi_A^0
       \end{array}
      \right)$
 & $\delta
    = \left(
       \begin{array}{cc}
	\frac{\delta^+}{\sqrt{2}}
	 & \delta^{++}\\
	\delta^0
	 & -\frac{\delta^+}{\sqrt{2}}
       \end{array}
      \right)$
 & $\Delta_A
    = \left(
       \begin{array}{cc}
	\frac{\Delta_A^+}{\sqrt{2}}
	 & \Delta_A^{++}\\
	\Delta_A^0
	 & -\frac{\Delta_A^+}{\sqrt{2}}
       \end{array}
      \right)$
\\\hline\hline
 $A_4$
 & ${\bf 1}$
 & ${\bf 1}^\prime$
 & ${\bf 1}^\dprime$
 & ${\bf 3}$
 & ${\bf 3}$
 & ${\bf 1}$
 & ${\bf 3}$
\\\hline
 $SU(2)_L$
 & ${\bf 1}$
 & ${\bf 1}$
 & ${\bf 1}$
 & ${\bf 2}$
 & ${\bf 2}$
 & ${\bf 3}$
 & ${\bf 3}$
\\\hline
 $U(1)_Y$
 & $-2$
 & $-2$
 & $-2$
 & $-1$
 & $1$
 & $2$
 & $2$
\end{tabular}
\caption{
 The leptons and the Higgs bosons in the A4HTM\@.
 The subscript $A = x, y, z$ denotes
the index for ${\bf 3}$ of $A_4$;
 for example,
$(\Psi_{xL}, \Psi_{yL}, \Psi_{zL})$ belongs to ${\bf 3}$
while each $\Psi_{AL}$ are $SU(2)_L$-doublet fields.
}
\label{tab:particle}
\end{table}

 The particle contents in the A4HTM
are listed in Table~\ref{tab:particle}.
 Singlet charged fermions
$\psi_{1R}^-$, $\psi_{2R}^-$, and $\psi_{3R}^-$
belong to ${\bf 1}$, ${\bf 1^\prime}$, and ${\bf 1^\dprime}$,
respectively.
 Doublet fermions,
$\Psi_{xL}$, $\Psi_{yL}$, and $\Psi_{zL}$
are members of ${\bf 3}$.
 A ${\bf 3}$-representation is composed of
Higgs doublets,
$\Phi_x$, $\Phi_y$, and $\Phi_z$.
 A triplet field $\delta$ of Higgs bosons
is of ${\bf 1}$.
 Three Higgs triplets,
$\Delta_x$, $\Delta_y$, and $\Delta_z$
construct a ${\bf 3}$-representation.
 Thus,
the A4HTM is a four-Higgs-Triplet-Model
and a three-Higgs-Doublet-Model
(we may introduce an extra doublet boson for quarks).
 Other versions of $A_4$-symmetric HTM
can be seen in \cite{Ma:2004zv,Hirsch:2005mc}
which have six triplet fields of
${\bf 1}$, ${\bf 1}^\prime$, ${\bf 1}^\dprime$,
and ${\bf 3}$.
 The calculations in this section
are almost identical to those for the model
in \cite{Altarelli:2005yp}
where $A_4$ is broken by vev's
of gauge singlet scalars (so-called flavon).
 The vev's of seven Higgs fields in the A4HTM
are taken as follows:
\begin{eqnarray}
&&
\langle \phi_x^0 \rangle
= \langle \phi_y^0 \rangle
= \langle \phi_z^0 \rangle
= \frac{v}{\sqrt{6}},
\label{eq:vev2}\\
&&
\langle \delta^0 \rangle
= \frac{v_\delta}{\sqrt{2}}, \quad
\langle \Delta_x^0 \rangle
= \frac{v_\Delta}{\sqrt{2}}, \quad
\langle \Delta_y^0 \rangle
= \langle \Delta_z^0 \rangle
= 0 ,
\label{eq:vev3}
\end{eqnarray}
where $v=246\,\GeV$.
 Similarly to the HTM,
triplet vev's $v_\delta$ and $v_\Delta$
should be generated
by explicit breaking terms of the lepton number conservation
because spontaneous breaking of it~\cite{Gelmini:1980re}
brings undesired Nambu-Goldston bosons (so-called Majoron).
 The triplet vev's
(and explicit breaking parameters for them)
are taken to be real positive
by using two phase degrees of freedom
of $\delta$ and $(\Delta_x, \Delta_y, \Delta_z)$.
 Note that
triplet vev's are constrained as
$v^\prime \equiv \sqrt{ v_\delta^2 +v_\Delta^2 } < 3\,\GeV$
by $\rho_0 = 1.0004^{+0.0027}_{-0.0007}$ at $2\sigma$~CL
(page 137 of \cite{Amsler:2008zzb}).
 Since the alignment eq.~(\ref{eq:vev2}) is invariant
for acting $T$ which satisfies $T^3=1$,
the A4HTM has an approximate $Z_3$ symmetry
which is broken only by a small $v_\Delta$.
 The Yukawa terms for doublet Higgs bosons are expressed as
\begin{eqnarray}
{\mathcal L}_{\text{d-Yukawa}}
=
 y_1 \bigl(\, \overline{\Psi_L}\, \Phi \bigr)_{\bf 1} \psi_{1R}
 + y_2 \bigl(\, \overline{\Psi_L}\, \Phi \bigr)_{{\bf 1}^\dprime} \psi_{2R}
 + y_3 \bigl(\, \overline{\Psi_L}\, \Phi \bigr)_{{\bf 1}^\prime} \psi_{3R}
 + \text{h.c.}
\end{eqnarray}
 The expressions $({\bf 3} \ {\bf 3})_{\bf 1}$ etc.\ mean
the decompositions of ${\bf 3}\otimes{\bf 3} \to {\bf 1}$ etc.\
among
${\bf 3}\otimes{\bf 3}
= {\bf 1} \oplus {\bf 1}^\prime \oplus {\bf 1}^\dprime
\oplus {\bf 3}_s \oplus {\bf 3}_a$
(See Appendix~\ref{app:A4}).
 The flavor eigenstates of leptons%
\footnote{
If $\delta$ belongs to ${\bf 1}^\prime$ instead of ${\bf 1}$,
the names of lepton flavors in eq.~(\ref{eq:lepton})
are changed as
$(e, \mu, \tau) \to (\mu, \tau, e)$
in order to keep the structure of the neutrino mixing.
}
are given by
\begin{eqnarray}
&&
\begin{pmatrix}
 e_R\\
 \mu_R\\
 \tau_R
\end{pmatrix}
\equiv
 U_R^\dagger
 \begin{pmatrix}
  \psi_{1R}^-\\
  \psi_{2R}^-\\
  \psi_{3R}^-
 \end{pmatrix}, \quad
\begin{pmatrix}
 L_e\\
 L_\mu\\
 L_\tau
\end{pmatrix}
\equiv
 U_L^\dagger
 \begin{pmatrix}
  \Psi_{xL}\\
  \Psi_{yL}\\
  \Psi_{zL}
 \end{pmatrix}, \quad
L_\ell
\equiv
 \begin{pmatrix}
  \nu_{\ell L}\\
  \ell_L
 \end{pmatrix} ,
\label{eq:lepton}\\
&&
U_R
\equiv
 \begin{pmatrix}
  1 & 0 & 0\\
  0 & 1 & 0\\
  0 & 0 & -1
 \end{pmatrix}, \quad
U_L
\equiv
 \frac{1}{\sqrt{3}}
 \begin{pmatrix}
  1 &        1 & 1\\
  1 &   \omega & \omega^2\\
  1 & \omega^2 & \omega
 \end{pmatrix}
 \begin{pmatrix}
  1 & 0 & 0\\
  0 & 1 & 0\\
  0 & 0 & -1
 \end{pmatrix}.
\end{eqnarray}
 The masses of charged leptons are
\begin{eqnarray}
m_e \equiv \frac{1}{\sqrt{2}}\, v y_1 , \quad
m_\mu \equiv \frac{1}{\sqrt{2}}\, v y_2 , \quad
m_\tau \equiv \frac{1}{\sqrt{2}}\, v y_3 .
\end{eqnarray}
 It is worth to note that
$L_e$, $L_\mu$, and $L_\tau$ are eigenstates of $T$
for eigenvalues $1$, $\omega$, and $\omega^2$,
respectively.

 Neutrinos in the A4HTM are Majorana fermions.
 In general,
the mass matrix $M_\nu$ of Majorana neutrinos
in the flavor basis can be expressed as
\begin{eqnarray}
M_\nu
&=&
 U_\MNS^\ast\,
 \text{diag}
 ( m_1 e^{i \alpha_{12}}, m_2, m_3 e^{i \alpha_{32}} )\,
 U_\MNS^\dagger ,
\label{eq:Mnu}
\end{eqnarray}
where $m_i$ are real positive masses.
 The parameters $\alpha_{12}$ and $\alpha_{32}$
within $[0, 2\pi)$
are Majorana phases~\cite{Schechter:1980gr,Mphase}
which appear only for Majorana particles.
 The standard parametrization of
the Maki-Nakagawa-Sakata (MNS) matrix~\cite{Maki:1962mu},
$U_\MNS$, is
\begin{eqnarray}
U_\MNS
=
 \begin{pmatrix}
  1 & 0 & 0\\
  0 & c_{23} & s_{23}\\
  0 & -s_{23} & c_{23} 
 \end{pmatrix}
 \begin{pmatrix}
  c_{13} & 0 & s_{13}\, e^{-i\delta_D}\\
  0 & 1 & 0\\
  -s_{13}\, e^{i\delta_D} & 0 & c_{13} 
 \end{pmatrix}
 \begin{pmatrix}
  c_{12} & s_{12} & 0\\
  -s_{12} & c_{12} & 0\\
  0 & 0& 1 
 \end{pmatrix} ,
\label{eq:MNS}
\end{eqnarray}
where $c_{ij}$ and $s_{ij}$ stand for
$\cos{\theta_{ij}}$ and $\sin{\theta_{ij}}$,
respectively.
 Neutrino oscillation measurements%
~\cite{solar,atm,acc,Apollonio:2002gd,:2008ee} show
\begin{eqnarray}
&&
 \Delta m^2_{21} \simeq 7.6\times 10^{-5}\,\eV^2 , \quad
 |\Delta m^2_{31}| \simeq 2.4\times 10^{-3}\,\eV^2 ,
\label{eq:dm2}\\
&&
 \sin^2{2\theta_{23}} \simeq 1 , \quad
 \sin^2{2\theta_{12}} \simeq 0.87 , \quad
 \sin^2{2\theta_{13}} \lesssim 0.14 ,
\label{eq:theta}
\end{eqnarray}
where $\Delta m^2_{ij} \equiv m_i^2 - m_j^2$.

 In the A4HTM,
neutrino masses are generated
by the Yukawa terms of triplet Higgs bosons:
\begin{eqnarray}
{\mathcal L}_{\text{t-Yukawa}}
=
 h_\delta
 \left[\,
  \overline{(\Psi_{L})_\alpha^c}\, (\Psi_{L})_\beta
 \right]_{\bf 1}
 ( i\sigma^2 \delta )_{\alpha\beta}
 +
 h_\Delta
 \Bigl(
  \Bigl(\,
   \overline{(\Psi_{L})_\alpha^c}\, (\Psi_{L})_\beta
  \Bigl)_{{\bf 3}_s}
  ( i\sigma^2 \Delta )_{\alpha\beta}
 \Bigr)_{\bf 1}
 + \text{h.c.} ,
\label{eq:tyukawa}
\end{eqnarray}
where $\alpha$ and $\beta$ stand for the $SU(2)_L$ index,
$\sigma^i$ are the Pauli matrices,
and the superscript $c$ means the charge conjugation.
 Without loss of generality,
$h_\delta$ can be taken as a real parameter
by the redefinition of the phase of $\Psi_L$.
 The decomposition indicated by
$\left[ {\bf 3}\ {\bf 3} \right]_{\bf 1}$
is the one which depends on the representation
(${\bf 1}$ or ${\bf 1}^\prime$ or ${\bf 1}^\dprime$)
of $\delta$.
 By using $U_L$ and triplet vev's for eq.~(\ref{eq:tyukawa}),
the mass matrix $M_\nu$ of neutrinos is obtained.
 The mass matrix is expressed in the form
of eq.~(\ref{eq:Mnu}) with
\begin{eqnarray}
&&
m_1 e^{i\alpha_{12}}
= h_\delta v_\delta + h_\Delta v_\Delta , \quad
m_2
= h_\delta v_\delta ,\quad
m_3 e^{i\alpha_{32}}
= - h_\delta v_\delta + h_\Delta v_\Delta ,
\\
&&
U_\MNS
= U_\TBM
\equiv
 \begin{pmatrix}
  \sqrt{\frac{2}{\,3\,}}
   & \frac{1}{\sqrt{3}}
   & 0\\
  -\frac{1}{\sqrt{6}}
   & \frac{1}{\sqrt{3}}
   & \frac{1}{\sqrt{2}}\\
  \frac{1}{\sqrt{6}}
   & -\frac{1}{\sqrt{3}}
   & \frac{1}{\sqrt{2}}
 \end{pmatrix} .
\end{eqnarray}
 $U_\TBM$ is the matrix of so-called tri-bimaximal mixing%
~\cite{Harrison:2002er}
which agrees with eq.~(\ref{eq:theta}).
 It is an attractive feature of the $A_4$ symmetry
that such a nontrivial mixing matrix can be given by a simple choice
of the vev's in eq.~(\ref{eq:vev2}) and (\ref{eq:vev3}).
 Two combinations of parameters are determined
by eq.~(\ref{eq:dm2}) as
\begin{eqnarray}
| h_\Delta | v_\Delta
&=&
 \frac{1}{\sqrt{2}}
 \sqrt{ \Delta m^2_{31} - 2 \Delta m^2_{21} }
\simeq 3.4\times 10^{-2}\, \eV ,
\label{eq:hDvD}\\
h_\delta v_\delta \cos\varphi_\Delta
&=&
 -
 \frac{ \Delta m^2_{31} }
      { 2\sqrt{2} \sqrt{ \Delta m^2_{31} - 2 \Delta m^2_{21} } }
\simeq - 1.8\times 10^{-2}\, \eV ,
\label{eq:hdvd}
\end{eqnarray}
where $\varphi_\Delta \equiv \text{arg}(h_\Delta)$.
 It is apparent in eq.~(\ref{eq:hDvD}) that
the A4HTM predicts $\Delta m^2_{31} > 0$.
 Then,
$m_i$ are given%
\footnote{
 Arbitrary $m_i$ can be obtained
if we introduce also $\delta_2$ of ${\bf 1}^\prime$
and $\delta_3$ of ${\bf 1}^\dprime$
with a condition
$h_{\delta 2} v_{\delta 2} = h_{\delta 3} v_{\delta 3}$
for their Yukawa couplings and vev's~\cite{Ma:2004zv}.
}
by
\begin{eqnarray}
m_1^2
&=&
 \left\{
  \frac{ 1 }
       { 8 (1-2r) \cos^2\varphi_\Delta }
  - r
 \right\} \Delta m^2_{31}
\geq
 (0.016 \eV)^2 ,
\\
m_2^2
&=&
 \frac{ \Delta m^2_{31} }
       { 8 (1-2r) \cos^2\varphi_\Delta }
\geq
 (0.018 \eV)^2 ,
\\
m_3^2
&=&
 \left\{
  \frac{ 1 }
       { 8 (1-2r) \cos^2\varphi_\Delta }
  + 1 - r
 \right\} \Delta m^2_{31}
\geq
 (0.051 \eV)^2 ,
\end{eqnarray}
where $r \equiv \Delta m^2_{21} / \Delta m^2_{31}$.
 Majorana phases are
\begin{eqnarray}
\tan{\alpha_{12}}
&=&
 -
 \frac{ ( 1 - 2 r ) \sin{2\varphi_\Delta} }
      { 1 - 2 (1 - 2r)  \cos^2{\varphi_\Delta} }, \\
\tan{\alpha_{32}}
&=&
 \frac{ ( 1 - 2r ) \sin{2\varphi_\Delta} }
      { 1 + 2 (1 - 2r)  \cos^2{\varphi_\Delta} }, \quad
 \cos\alpha_{32} < 0 .
\end{eqnarray}
 Numerically,
$|\alpha_{32} - \pi| \lesssim 0.16 \pi$.
 The effective mass $(M_\nu)_{ee}$
for the neutrinoless double beta decay
(See \cite{Avignone:2007fu} for a review)
is expressed as
\begin{eqnarray}
| (M_\nu)_{ee} |^2
=
 \left(
  \frac{1}{ 8 (1 - 2r) \cos^2{\varphi_\Delta} }
  - \frac{1+4r}{9}
 \right)
 \Delta m^2_{31}
\geq
 ( 0.0045\,\eV )^2 .
\end{eqnarray}

%%%%%%%%  sect: Higgs sector  %%%%%%
\section{Higgs Sector}\label{sect:Higgs}
 It is necessary to take mass eigenstates of the Higgs bosons
in order to consider their phenomenology
which is our purpose in this article.
 The mass eigenstates can be obtained
from the Higgs potential shown in the next subsection.

%%%%%  subsect: Higgs Potential  %%%%
\subsection{Higgs Potential}
 Let us first remind that
an expression~\cite{Chun:2003ej} of the Higgs potential
in the HTM without the $A_4$ symmetry is
\begin{eqnarray}
V_{\text{HTM}}
&=&
 - m^2(\Phi^\dagger\Phi)
 + \lambda_1 (\Phi^\dagger\Phi)^2
 + M^2 {\rm Tr}(\Delta^\dagger\Delta)
 + \lambda_2[{\rm Tr}(\Delta^\dagger\Delta)]^2
 + \lambda_3{\rm Det} (\Delta^\dagger\Delta)
\nonumber \\
&&\hspace*{-2mm}
{}+\lambda_4 (\Phi^\dagger\Phi) {\rm Tr}(\Delta^\dagger\Delta)
  +\lambda_5 (\Phi^\dagger\sigma^i\Phi)
             {\rm Tr}(\Delta^\dagger\sigma^i\Delta)
  +\left(
    \frac{1}{\sqrt{2}} \mu(\Phi^Ti\sigma^2\Delta^\dagger\Phi)
    + \text{h.c.}
   \right) ,
\label{eq:VHTM}
\end{eqnarray}
where $\Phi$ and $\Delta$ are doublet and triplet
Higgs bosons, respectively.
 Using these notations of coupling constants as the reference,
we construct the $A_4$-symmetric potential
for the A4HTM as
\begin{eqnarray}
V_{\text{A4HTM}}
&\equiv&
 V_m
 + V_1 + V_2 + V_3 + V_4 + V_5
 + V_\mu ,
\\
V_m
&\equiv&
 -
 m_\Phi^2\, (\Phi^\dagger \Phi)_{\bf 1}
 +
 M_\delta^2\, \tr( \delta^\dagger \delta )
 +
 M_\Delta^2\, \tr( \Delta^\dagger \Delta)_{\bf 1} ,
\\
V_4
&\equiv&
 \lambda_{4 \delta}\,
 ( \Phi^\dagger \Phi )_{\bf 1}\,
 \tr\bigl( \delta^\dagger \delta \bigr)
 +
 \lambda_{4 \Delta}\,
 ( \Phi^\dagger \Phi )_{\bf 1}\,
 \tr(\Delta^\dagger \Delta )_{\bf 1}
\nonumber\\
&&\hspace*{0mm}
{}
 +
 \left\{
  \lambda_{4 \Delta p}^\prime\,
  ( \Phi^\dagger \Phi )_{{\bf 1}^\dprime}\,
  \tr( \Delta^\dagger \Delta )_{{\bf 1}^\prime}
  +
  \text{h.c.}
 \right\}
\nonumber\\
&&\hspace*{0mm}
{}
 +
 \lambda_{4 \Delta ss}\,
 ( \Phi^\dagger \Phi )_{{\bf 3}_s}\,
 \tr( \Delta^\dagger \Delta )_{{\bf 3}_s}
 +
 \lambda_{4 \Delta aa}\,
 ( \Phi^\dagger \Phi )_{{\bf 3}_a}\,
 \tr( \Delta^\dagger \Delta )_{{\bf 3}_a}
\nonumber\\
&&\hspace*{0mm}
{}
 +
 i\lambda_{4 \Delta sa}\,
 ( \Phi^\dagger \Phi )_{{\bf 3}_s}\,
 \tr( \Delta^\dagger \Delta )_{{\bf 3}_a}
 +
 i\lambda_{4 \Delta as}\,
 ( \Phi^\dagger \Phi )_{{\bf 3}_a}\,
 \tr( \Delta^\dagger \Delta )_{{\bf 3}_s}
\nonumber\\
&&\hspace*{0mm}
{}
 +
 \left\{
  \lambda_{4s}^\prime\,
  \delta_{\beta\alpha}^\ast\,
  \bigl[
   \Delta_{\beta\alpha}\,
   ( \Phi^\dagger \Phi )_{{\bf 3}_s}
  \bigr]_{\bf 1}
  +
  \lambda_{4a}^\prime\,
  \delta_{\beta\alpha}^\ast\,
  \bigl[
   \Delta_{\beta\alpha}\,
   ( \Phi^\dagger \Phi )_{{\bf 3}_a}
  \bigr]_{\bf 1}
  +
  \text{h.c.}
 \right\},
\\
V_5
&\equiv&
 \lambda_{5 \delta}\,
 ( \Phi^\dagger \sigma^i \Phi )_{\bf 1}\,
 \tr( \delta^\dagger \sigma^i \delta )
 +
 \lambda_{5 \Delta}\,
 ( \Phi^\dagger \sigma^i \Phi )_{\bf 1}\,
 \tr(\Delta^\dagger \sigma^i \Delta )_{\bf 1}
\nonumber\\
&&\hspace*{0mm}
{}
 +
 \left\{
  \lambda_{5 \Delta p}^\prime\,
  ( \Phi^\dagger \sigma^i \Phi )_{{\bf 1}^\dprime}\,
  \tr( \Delta^\dagger \sigma^i \Delta )_{{\bf 1}^\prime}
  +
  \text{h.c.}
 \right\}
\nonumber\\
&&\hspace*{0mm}
{}
 +
 \lambda_{5 \Delta ss}\,
 ( \Phi^\dagger \sigma^i \Phi )_{{\bf 3}_s}\,
 \tr( \Delta^\dagger \sigma^i \Delta )_{{\bf 3}_s}
 +
 \lambda_{5 \Delta aa}\,
 ( \Phi^\dagger \sigma^i \Phi )_{{\bf 3}_a}\,
 \tr( \Delta^\dagger \sigma^i \Delta )_{{\bf 3}_a}
\nonumber\\
&&\hspace*{0mm}
{}
 +
 i \lambda_{5 \Delta sa}\,
 ( \Phi^\dagger \sigma^i \Phi )_{{\bf 3}_s}\,
 \tr( \Delta^\dagger \sigma^i \Delta )_{{\bf 3}_a}
 +
 i \lambda_{5 \Delta as}\,
 ( \Phi^\dagger \sigma^i \Phi )_{{\bf 3}_a}\,
 \tr( \Delta^\dagger \sigma^i \Delta )_{{\bf 3}_s}
\nonumber\\
&&\hspace*{0mm}
{}
 +
 \bigl\{
  \lambda_{5s}^\prime\,
  ( \delta^\dagger \sigma^i )_{\alpha\beta}\,
  \bigl[
   \Delta_{\beta\alpha}\,
   ( \Phi^\dagger \sigma^i \Phi )_{{\bf 3}_s}
  \bigr]_{\bf 1}
\nonumber\\
&&\hspace*{20mm}
{}
  +
  \lambda_{5a}^\prime\,
  ( \delta^\dagger \sigma^i )_{\alpha\beta}\,
  \bigl[
   \Delta_{\beta\alpha}\,
   ( \Phi^\dagger \sigma^i \Phi )_{{\bf 3}_a}
  \bigr]_{\bf 1}
  +
  \text{h.c.}
 \bigr\},
\end{eqnarray}
where coupling constants $\lambda^\prime$ have complex values
while $\lambda$'s are real%
\footnote
{
 One may rewrite $V_5$ with
$(\Phi_A^\dagger \sigma^i \Phi_B) \tr(\Delta_C^\dagger \sigma^i \Delta_D)
= 2 \Phi_A^\dagger \Delta_D \Delta_C^\dagger \Phi_B
- (\Phi_A^\dagger \Phi_B) \tr(\Delta_C^\dagger \Delta_D)$.
}
.
 The subscripts $\alpha$ and $\beta$
stand for the indices of $SU(2)_L$.
 Main parts of the squared mass matrices for triplet fields
are induced by $V_m$, $V_4$, and $V_5$,
which give $v^2 \Delta^{--}_x \Delta^{++}_x$ etc.
 Contributions from $V_2$ and $V_3$ can be ignored
because they are suppressed by small triplet vev's.
 The expressions of $V_1$, $V_2$, $V_3$, and $V_\mu$
are presented in Appendix~\ref{app:V}.
 Linear terms of triplet fields exist not only in $V_\mu$
but also in $V_4$ and $V_5$,
which affect vacuum conditions for triplet vev's.
 Actually,
the democratic alignment of doublet vev's in eq.~(\ref{eq:vev2})
results in the democratic one for triplet ${\bf 3}$ also,
which conflicts with eq.~(\ref{eq:vev3}).
 Some solutions on the alignment problem
were discussed in \cite{Altarelli:2005yp}.
 We may simply assume
\begin{eqnarray}
v_\delta\, \text{Re}(\lambda_{4s}^\prime + \lambda_{5s}^\prime )
 + v_\Delta ( \lambda_{4\Delta ss} + \lambda_{5 \Delta ss} )
&=& 0 ,
\label{eq:assmpt}
\end{eqnarray}
and use $\tilde{V}_\mu$ with the soft breaking of $A_4$
instead of the $A_4$-symmetric $V_\mu$;
 for example,
\begin{eqnarray}
&&
\tilde{V}_\mu
=
 \frac{1}{\sqrt{2}} \mu_\delta
 \bigl[ \Phi_\alpha \Phi_\beta \bigr]_{\bf 1}
 (i\sigma^2 \delta^\dagger)_{\alpha\beta}
 +
 \frac{1}{\sqrt{2}} \mu_{\Delta_x}
 ( 2 \Phi_{y\alpha} \Phi_{z\beta} )
 (i\sigma^2 \Delta_x^\dagger)_{\alpha\beta}
 +
 \text{h.c.},
\end{eqnarray}
where $\mu_{\Delta_x}$ breaks softly
the lepton number conservation and the $A_4$ symmetry%
\footnote
{
 If representation of $\delta$ is ${\bf 1}^\prime$,
also $\mu_\delta$ must break $A_4$
because $(\Phi_\alpha \Phi_\beta)_{{\bf 1}^\dprime}$
does not contain $v^2$ term.
}.
 Redefinitions of phases of
$\delta$ and ($\Delta_x$, $\Delta_y$, $\Delta_z$)
enable us to make $\mu_\delta$ and $\mu_{\Delta x}$
real positive parameters.
 Ignoring corrections due to small triplet vev's,
we can have
\begin{eqnarray}
v
&=&
 \sqrt{6} \langle \phi_x^0 \rangle
\simeq \sqrt{6} \langle \phi_y^0 \rangle
\simeq \sqrt{6} \langle \phi_z^0 \rangle
\simeq
 \frac{ \sqrt{3}\, m_\Phi }
      { \sqrt{ 3\lambda_1 + 4\lambda_{1ss} } } ,
\\
\begin{pmatrix}
 v_\delta\\
 v_\Delta
\end{pmatrix}
&\simeq&
 \begin{pmatrix}
  6M_\delta^2 + 3 \lambda_{\delta 45p} v^2
   & 2 \text{Re}(\lambda_{s 45p}^\prime) v^2\\
  2 \text{Re}(\lambda_{s 45p}^\prime) v^2
   & 6M_\Delta^2 + 3 \lambda_{\Delta 45p} v^2
 \end{pmatrix}^{-1}
 \begin{pmatrix}
  3 v^2 \mu_\delta\\
  2 v^2 \mu_{\Delta_x}
 \end{pmatrix} ,
\\
\langle \Delta_y^0 \rangle
&=&
 \langle \Delta_z^0 \rangle
= 0 ,
\end{eqnarray}
where $\lambda_{45p}$'s
are defined by $\lambda_4 + \lambda_5$
for each subscripts;
for example,
$\lambda_{\delta 45p} \equiv \lambda_{4\delta} + \lambda_{5\delta}$.
 Small triplet vev's may be also the origin of
the small deviation from the tri-bimaximal mixing
(small $\theta_{13}$).
 In the following parts of this article,
we just use the vacuum alignment
in eq.~(\ref{eq:vev2}) and (\ref{eq:vev3})
ignoring how to achieve them.

%%%%%%  H++  %%%%%%%%%%%%%%
\subsection{Mass Eigenstates of Triplet Higgs Bosons}
 Ignoring small contributions from triplet vev's,
the squared mass matrix of doubly charged Higgs bosons
is obtained from $V_m + V_4 + V_5$ as
\begin{eqnarray}
&&
\begin{pmatrix}
 \Delta_x^{--} \
 \Delta_y^{--} \
 \Delta_z^{--} \
 \delta^{--}
\end{pmatrix}
\nonumber\\
&&\hspace*{10mm}
\times
\begin{pmatrix}
 M_{\Delta 45m}^2
  & \bigl[ M_{\pm\pm}^2 \bigr]_{21}^\ast
  & \bigl[ M_{\pm\pm}^2 \bigr]_{21}
  & \frac{1}{\,3\,} v^2 ( \lambda_{s 45m}^\prime )^\ast
\\
 \bigl[ M_{\pm\pm}^2 \bigr]_{21}
  & M_{\Delta 45m}^2
  & \bigl[ M_{\pm\pm}^2 \bigr]_{21}^\ast
  & \frac{1}{\,3\,} v^2 ( \lambda_{s 45m}^\prime )^\ast
\\
 \bigl[ M_{\pm\pm}^2 \bigr]_{21}^\ast
  & \bigl[ M_{\pm\pm}^2 \bigr]_{21}
  & M_{\Delta 45m}^2
  & \frac{1}{\,3\,} v^2 ( \lambda_{s 45m}^\prime )^\ast
\\
 \frac{1}{\,3\,} v^2 \lambda_{s 45m}^\prime
  & \frac{1}{\,3\,} v^2 \lambda_{s 45m}^\prime
  & \frac{1}{\,3\,} v^2 \lambda_{s 45m}^\prime
  & M_{\delta 45m}^2
\end{pmatrix}
\begin{pmatrix}
 \Delta_x^{++}\\
 \Delta_y^{++}\\
 \Delta_z^{++}\\
 \delta^{++}
\end{pmatrix},
\\
&&
\bigl[ M_{\pm\pm}^2 \bigr]_{21}
\equiv
 \frac{1}{\,3\,} v^2
 \left(
  \lambda_{\Delta ss 45m} + i \lambda_{\Delta sa 45m}
 \right),
\\
&&
M_{\delta 45m}^2
\equiv
 M_\delta^2
 +
 \frac{1}{\,2\,} v^2 \lambda_{\delta 45m}, \quad
M_{\Delta 45m}^2
\equiv
 M_\Delta^2
 +
 \frac{1}{\,2\,} v^2 \lambda_{\Delta 45m},
\label{eq:MD45m}
\end{eqnarray}
where $\lambda_{45m}$'s are defined by
$\lambda_4 - \lambda_5$ for each subscripts;
 for example,
$\lambda_{\Delta ss 45m}
\equiv \lambda_{4 \Delta ss} - \lambda_{5 \Delta ss}$.
 Then,
the mass eigenstates of doubly charged Higgs bosons
are given by
\begin{eqnarray}
\begin{pmatrix}
 H_1^{++}\\
 H_2^{++}\\
 H_3^{++}\\
 H_4^{++}
\end{pmatrix}
&=&
 \frac{1}{\sqrt{3}}
 \begin{pmatrix}
  1
   & 0
   & 0
   & 0\\
  0
   & 1
   & 0
   & 0\\
  0
   & 0
   & \cos\theta_{\pm\pm}
   & \sin\theta_{\pm\pm}\\
  0
   & 0
   & - \sin\theta_{\pm\pm}
   & \cos\theta_{\pm\pm}
 \end{pmatrix}
 \begin{pmatrix}
  1
   & \omega
   & \omega^2
   & 0\\
  1
   & \omega^2
   & \omega
   & 0\\
  1
   & 1
   & 1
   & 0\\
  0
   & 0
   & 0
   & \sqrt{3}\,e^{-i\, \text{arg}(\lambda_{s 45m}^\prime)}
 \end{pmatrix}
% U_{\pm\pm}
 \begin{pmatrix}
  \Delta_x^{++}\\
  \Delta_y^{++}\\
  \Delta_z^{++}\\
  \delta^{++}
 \end{pmatrix} ,
\label{eq:Upp}\\
\tan{2\theta_{\pm\pm}}
&\equiv&
 \frac{ 2\sqrt{3}\, |\lambda_{s 45m}^\prime|\, v^2 }
      {
       3 M_{\Delta 45m}^2
       - 3 M_{\delta 45m}^2
       + 2 \lambda_{\Delta ss 45m} v^2
      } ,
\label{eq:tanpp}
\end{eqnarray}
where $0 \leq \theta_{\pm\pm} \leq \pi/4$
for negative values ($\leq 0$) of the denominator of eq.~(\ref{eq:tanpp})
and $\pi/4 < \theta_{\pm\pm} \leq \pi/2$
for positive values ($> 0$).
 It is understood
by the approximate $Z_3$ symmetry of the A4HTM
that $\delta$ is mixed with
$\Delta_\xi \equiv (\Delta_x + \Delta_y + \Delta_z)/\sqrt{3}$
for which acting $T$ gives $1$ as the eigenvalue%
\footnote{
 If $\delta$ belongs to ${\bf 1}^\prime$,
the field is mixed with
$\Delta_\eta \equiv
(\Delta_x + \omega^2 \Delta_y + \omega \Delta_z)/\sqrt{3}$
which is an eigenstate of $T$
for an eigenvalue $\omega$.
 There will be no difficulty to obtain
mass eigenstates of Higgs bosons
even in the model of \cite{Ma:2004zv}
where $\delta_2$ of ${\bf 1}^\prime$
and $\delta_3$ of ${\bf 1}^\dprime$
are also introduced.
}.
 The masses $m_{H_i^{\pm\pm}}$ of $H_i^{\pm\pm}$ are
\begin{eqnarray}
m_{H_1^{\pm\pm}}^2
&=&
 M_{\Delta 45m}^2
 - \frac{1}{\,3\,}
   \lambda_{\Delta ss 45m} v^2
 + \frac{1}{\sqrt{3}}
   \lambda_{\Delta sa 45m} v^2 ,
\label{eq:m1pp}
\\
m_{H_2^{\pm\pm}}^2
&=&
 M_{\Delta 45m}^2
 - \frac{1}{\,3\,}
   \lambda_{\Delta ss 45m} v^2
 - \frac{1}{\sqrt{3}}
   \lambda_{\Delta sa 45m} v^2 ,
\label{eq:m2pp}
\\
m_{H_3^{\pm\pm}}^2
&=&
 \frac{1}{\,6\,}
 \left(
  3 M_{\delta 45m}^2
  + 3 M_{\Delta 45m}^2
  + 2 \lambda_{\Delta ss 45m} v^2
  - 3 \Delta m_{\pm\pm}^2
 \right) ,
\label{eq:m3pp}
\\
m_{H_4^{\pm\pm}}^2
&=&
 \frac{1}{\,6\,}
 \left(
  3 M_{\delta 45m}^2
  + 3 M_{\Delta 45m}^2
  + 2 \lambda_{\Delta ss 45m} v^2
  + 3 \Delta m_{\pm\pm}^2
 \right) ,
\label{eq:m4pp}
\\
3 \Delta m_{\pm\pm}^2
&\equiv&
 \left\{
  12 | \lambda_{s 45m}^\prime |^2 v^4
  + \left(
     3 M_{\Delta 45m}^2
     - 3 M_{\delta 45m}^2
     + 2 \lambda_{\Delta ss 45m} v^2
    \right)^2
 \right\}^{\frac{1}{\,2\,}} .
\label{eq:dmpp}
\end{eqnarray}
 Note that $m_{H_3^{\pm\pm}} \leq m_{H_4^{\pm\pm}}$
as the definition.
 These masses $m_{H_i^{\pm\pm}}$ can be different enough from each other
while the constraint from $\rho$-parameter
does not prefer large mass differences between
$H_i^{\pm\pm}$ and their triplet-like partners
($H_{Ti}^\pm$, $H_{Ti}^0$, and $A_{Ti}^0$).

 Decays of $H^{\pm\pm}_i$
into same-signed charged leptons in the flavor basis
are governed by the following couplings $h_{i\pm\pm}$
for $(h_{i\pm\pm})_{\ell\ell^\prime}
H^{++} \overline{(\ell_L)^c} \ell_L^\prime$:
\begin{eqnarray}
h_{1\pm\pm}
&=&
 \frac{1}{\sqrt{3}}\, h_\Delta
 \begin{pmatrix}
   0 & -1 & 0\\
  -1 &  0 & 0\\
   0 &  0 & 2
 \end{pmatrix} ,
\label{eq:h1pp}\\
h_{2\pm\pm}
&=&
 \frac{1}{\sqrt{3}}\, h_\Delta
 \begin{pmatrix}
  0 & 0 & 1\\
  0 & 2 & 0\\
  1 & 0 & 0
 \end{pmatrix} ,
\label{eq:h2pp}\\
h_{3\pm\pm}
&=&
 \frac{1}{\sqrt{3}}\,
 h_\Delta \cos\theta_{\pm\pm}
 \begin{pmatrix}
  2 & 0 & 0\\
  0 & 0 & 1\\
  0 & 1 & 0
 \end{pmatrix}
 +
 \tilde{h}_\delta
 \sin\theta_{\pm\pm}
 \begin{pmatrix}
  1 &  0 & 0\\
  0 &  0 & -1\\
  0 & -1 & 0
 \end{pmatrix} ,
\label{eq:h3pp}\\
h_{4\pm\pm}
&=&
 -\frac{1}{\sqrt{3}}\,
 h_\Delta \sin\theta_{\pm\pm}
 \begin{pmatrix}
  2 & 0 & 0\\
  0 & 0 & 1\\
  0 & 1 & 0
 \end{pmatrix}
 +
 \tilde{h}_\delta
 \cos\theta_{\pm\pm}
 \begin{pmatrix}
  1 &  0 & 0\\
  0 &  0 & -1\\
  0 & -1 & 0
 \end{pmatrix} ,
\label{eq:h4pp}\\
\tilde{h}_\delta
&\equiv&
 h_\delta e^{i\, \text{arg}(\lambda_{s 45m}^\prime)} .
\label{eq:dmpp}
\end{eqnarray}
 The zeros in $h_{i\pm\pm}$
can be understood easily by the eigenvalues of $T$
for eigenstates $H_i^{\pm\pm}$ and leptons
(Table~\ref{tab:t-ev});
for example,
$(h_{1\pm\pm})_{ee}$ must vanish approximately
because $\overline{(e_L)^c} e_L H_1^{++}$ is not invariant
for acting $T$.
 On the other hand,
the values of nonzero elements of $h_{i\pm\pm}$
are the consequence of the $A_4$ symmetry.

\begin{table}[t]
\begin{tabular}{c||c|c|c||c|c|c}
 {}
  & \ $e_L$ \
  & \ $\mu_L$ \
  & \ $\tau_L$ \
  & \ $H_1^{++}$ \
  & \ $H_2^{++}$ \
  & \ $H_3^{++}$, $H_4^{++}$ \
\\\hline\hline
 \ $T$ \
  & $1$
  & $\omega$
  & $\omega^2$
  & $\omega^2$
  & $\omega$
  & $1$
\end{tabular}
\caption{
 Eigenstates and eigenvalues of $T$.
}
\label{tab:t-ev}
\end{table}

 Next,
let us consider singly charged scalar fields also.
 We concentrate on
the four triplet-like singly charged scalar, $H^\pm_{Ti}$.
 The mixing between doublet and triplet bosons is ignored
because it is suppressed by small vev's of triplet fields%
\footnote
{
 The small vev can give a maximal mixing
for neutral bosons in a special case
but this does not happen for charged ones
of the interest in this article.
 Phenomenology for the case in the HTM
is shown in \cite{Akeroyd:2010je}.
}.
 Then,
we can diagonalize the squared mass matrix
for the singly charged ones of triplet fields
similarly to the case for doubly charged ones.
 The mixing matrix with an angle $\theta_\pm$,
masses $m_{H^\pm_{Ti}}$,
and couplings $h_{i\pm}$
for $\sqrt{2}\, (h_{i\pm})_{\ell\ell^\prime}
H^+_{Ti} \overline{(\nu_{\ell L})^c} \ell^\prime_L$
are given simply by setting $\lambda_5=0$
in eq.~(\ref{eq:Upp})-(\ref{eq:dmpp}).
 Note that $h_{3\pm}$ and $h_{4\pm}$ can be
different from $h_{3\pm\pm}$ and $h_{4\pm\pm}$
in the A4HTM, respectively,
while $h_\pm=h_{\pm\pm}$ in the HTM\@.
 This is because $\theta_\pm$ can be different
from $\theta_{\pm\pm}$ by the existence of $\lambda_5$
in principle.
 However,
$\theta_\pm \simeq \theta_{\pm\pm}$ seems preferred
because $m_{H_{Ti}^\pm} \simeq m_{H_i^{\pm\pm}}$
(namely, $|\lambda_5| \ll 1$)
is favored by the $\rho$ parameter.

 The mass eigenstates of the triplet-like
neutral Higgs bosons are shown in Appendix~\ref{app:neutral}
for completeness.

%%%%%%%%  sect:Phenomenology  %%%%%%%%%%%%%%
\section{Phenomenology of Higgs bosons}
\label{sect:pheno}

%%%%%%%%%%%%  Phenomenology of H++ %%%%%%%%%%%
 We assume that some of exotic Higgs bosons
are light enough to be detected in collider experiments
and to give sizable effects on some processes%
\footnote{
 Even if $M_\delta$ and $M_\Delta$ are very large,
the leptogenesis with the decays of triplet bosons~\cite{Ma:1998dx}
does not happen in this model
because their decays into $\Psi_L$ have
individual final states as we see in eq.~(\ref{eq:tyukawa}).
}.
 In this section,
we first list up exotic processes
which are possible with $H_i^{\pm\pm}$
and the triplet-like $H_{Ti}^\pm$.
 Then,
constraints from these processes are considered
in the next section.

\subsection{$H^{--} \to \ell \ell^\prime$}
%%%%%%%%  Table for H++  %%%%%%%%
\begin{table}[t]
\begin{tabular}{c||c|c|c}
 {}
 & \ $\BR(H^{--} \to \ell\ell^\prime)$ \
 & \ $\tlll$ \
 & others
\\
 {}
 & \ $ee:\mu\mu:\tau\tau:e\mu:e\tau:\mu\tau$ \
 &
 &
\\\hline\hline
 $H_1^{\pm\pm}$
 & $0:0:2:1:0:0$
 & none
 & 
\\\hline
 $H_2^{\pm\pm}$
 & $0:2:0:0:1:0$
 & \ $\tLeLmLmL$ \
 & 
\\\hline
 $H_3^{\pm\pm}$
 & $R^{\pm\pm}_3
    :0:0:0:0:
    1$ \hspace*{5mm}
 & \ $\tLmLeLeL$ \
 & \ $\overline{e_L} e_L \to \overline{e_L} e_L$ \
\\\hline
 $H_4^{\pm\pm}$
 & $R^{\pm\pm}_4
    :0:0:0:0:
    1$ \hspace*{5mm}
 & \ $\tLmLeLeL$ \
 & \ $\overline{e_L} e_L \to \overline{e_L} e_L$ \
\end{tabular}
\caption{
 Ratios of decays of $H^{\pm\pm}_i$
into a pair of same-signed charged leptons
in the A4HTM\@.
 Contributions of $H^{\pm\pm}_i$
to $\tlll$ at the tree level are also shown.
 Note that all of $H^{\pm\pm}_i$ does not
contribute to $\meee$ and $\llg$
at the tree and one loop level, respectively.
 The Bhabha scattering can be affected
by $H_3^{\pm\pm}$ and $H_4^{\pm\pm}$.
}
\label{tab:Hpp}
\end{table}

 The ratios of the branching ratios
$\BR_{\ell\ell^\prime} \equiv \BR( H^{--} \to \ell \ell^\prime )$
are shown in Table~\ref{tab:Hpp}.
 We used
\begin{eqnarray}
R^{\pm\pm}_3
\equiv
 \frac{ |2h_\Delta c_{\pm\pm} + \sqrt{3}\, \tilde{h}_\delta s_{\pm\pm}|^2 }
      { 2 |h_\Delta c_{\pm\pm} - \sqrt{3}\, \tilde{h}_\delta s_{\pm\pm}|^2 } ,
\quad
%\\
%
%
%
R^{\pm\pm}_4
\equiv
 \frac{ |2h_\Delta s_{\pm\pm} - \sqrt{3}\, \tilde{h}_\delta c_{\pm\pm}|^2 }
{ 2 |h_\Delta s_{\pm\pm} + \sqrt{3}\, \tilde{h}_\delta c_{\pm\pm}|^2 } ,
\end{eqnarray}
where $c_{\pm\pm} \equiv \cos\theta_{\pm\pm}$
and $s_{\pm\pm} \equiv \sin\theta_{\pm\pm}$.
 It is clear that each of $H_i^{\pm\pm}$ has
only two decay modes into a pair of same-signed charged leptons.
 For a simple case with $\tan{2\theta_{\pm\pm}}=0$,
one of $R_3^{\pm\pm}$ and $R_4^{\pm\pm}$ becomes $2$
while the other is $1/2$.
 Then,
the $H_i^{\pm\pm}$ which gives $\BR_{ee}/\BR_{\mu\tau} = 1/2$
can be identified as the $\delta^{\pm\pm}$ boson%
\footnote{
 Decays of $\delta^{\pm\pm}$
of ${\bf 1}^\prime$ or ${\bf 1}^\dprime$
also gives $\BR_{ee}/\BR_{\mu\tau} = 1/2$.
}.
 An interesting point is that
decays of $H_1^{\pm\pm}$ and $H_2^{\pm\pm}$ give
$\BR_{\mu\mu} \neq \BR_{\tau\tau}$
and $\BR_{e\mu} \neq \BR_{e\tau}$
in contrast with the case for the HTM
in which $\BR_{\mu\mu} \simeq \BR_{\tau\tau}$
and $\BR_{e\mu} \simeq \BR_{e\tau}$.
 If $m_{H_1^{\pm\pm}} = m_{H_2^{\pm\pm}}$
which is realized at $\lambda_{4\Delta sa} = \lambda_{5\Delta sa}$,
the sum of $\BR_{\ell\ell^\prime}$ of $H_1^{\pm\pm}$ and $H_2^{\pm\pm}$ gives
$\BR_{\mu\mu} = \BR_{\tau\tau}$
and $\BR_{e\mu} = \BR_{e\tau}$.

 If decays of $H_i^{\pm\pm}$ are
dominated by leptonic ones,
the A4HTM gives sharp predictions
for $\BR$'s themselves as
%\begin{align}
%\BR(H_1^{--} \to e \mu)
%&=
% \frac{1}{\,3\,} ,
%&
%\BR(H_1^{--} \to \tau\tau)
%&=
% \frac{2}{\,3\,} ,\\
%%
%%
%\BR(H_2^{--} \to \mu \mu)
%&=
% \frac{2}{\,3\,} ,
%&
%\BR(H_2^{--} \to e\tau)
%&=
% \frac{1}{\,3\,} ,\\
%%
%%
%\BR(H_3^{--} \to e e)
%&=
% \frac{ |2h_\Delta c_{\pm\pm} + \sqrt{3}\, \tilde{h}_\delta s_{\pm\pm}|^2 }
%      { 6 |h_\Delta|^2 c_{\pm\pm}^2 + 9 |\tilde{h}_\delta|^2 s_{\pm\pm}^2 } ,
%&
%\BR(H_3^{--} \to \mu\tau)
%&=
% 1 - \BR(H_3^{--} \to e e) ,\\
%%
%%
%\BR(H_4^{--} \to e e)
%&=
% \frac{ |2h_\Delta s_{\pm\pm} - \sqrt{3}\, \tilde{h}_\delta c_{\pm\pm}|^2 }
%      { 6 |h_\Delta|^2 s_{\pm\pm}^2 + 9 |\tilde{h}_\delta|^2 c_{\pm\pm}^2 } ,
%&
%\BR(H_4^{--} \to \mu\tau)
%&=
% 1 - \BR(H_4^{--} \to e e)
%\end{align}
\begin{eqnarray}
\BR(H_1^{--} \to e \mu)
&=&
 \frac{1}{\,3\,} ,\\
\BR(H_2^{--} \to \mu \mu)
&=&
 \frac{2}{\,3\,} ,\\
\BR(H_3^{--} \to e e)
&=&
 \frac{ |2h_\Delta c_{\pm\pm} + \sqrt{3}\, \tilde{h}_\delta s_{\pm\pm}|^2 }
      { 6 |h_\Delta|^2 c_{\pm\pm}^2 + 9 |\tilde{h}_\delta|^2 s_{\pm\pm}^2 } ,\\
\BR(H_4^{--} \to e e)
&=&
 \frac{ |2h_\Delta s_{\pm\pm} - \sqrt{3}\, \tilde{h}_\delta c_{\pm\pm}|^2 }
      { 6 |h_\Delta|^2 s_{\pm\pm}^2 + 9 |\tilde{h}_\delta|^2 c_{\pm\pm}^2 } ,
\end{eqnarray}
where modes involving $\tau$ are omitted.
 Especially,
$2/3$ for $\BR_{\mu\mu}$ is too large to be
reached in the HTM where $\BR_{\mu\mu} \lesssim 0.47$%
~\cite{Akeroyd:2007zv}.
 It is possible to have a large $\BR_{ee}$
which can not be explained by the HTM
where $\BR_{ee} \lesssim 0.49$;
for example,
the decay of $H_3^{\pm\pm}$ gives
$\BR_{ee} = 2/3$ for $\theta_{\pm\pm}=0$
and $\BR_{ee} = 1$ for
$h_\Delta c_{\pm\pm} = \sqrt{3}\, \tilde{h}_\delta s_{\pm\pm}$.
 Even if $\BR_{ee}$ turns out to be very small,
it does not result in a very small decay rate
of the neutrinoless double beta decay
(For the case in the HTM, see e.~g., \cite{Petcov:2009zr}
and references therein).
 Unfortunately,
it seems difficult to extract the information
on $\phi_\Delta \equiv \text{arg}(h_\Delta)$
from $\BR_{\ell\ell^\prime}$.

\subsection{$\tlll$ and others}
 The third column of Table~\ref{tab:Hpp}
shows possible $\tlll$ with $H_i^{\pm\pm}$ mediation
at the tree level.
 The most important point is that
$H_i^{\pm\pm}$ do not cause $\meee$ at the tree level,
for which the experimental constraint is very stringent
as $\BR(\meee) < 1.0\times 10^{-12}$~\cite{Bellgardt:1987du}.
 The radiative decays $\llg$ at one loop level with $H_i^{\pm\pm}$
are also forbidden.
 The eliminations of these lepton flavor violating decays
can be understood as the consequence
of the approximate $Z_3$ symmetry of the A4HTM\@.
 Therefore,
we can naturally expect signals of $\tlll$
in the future in collider experiments
(Super-KEKB~\cite{Akeroyd:2004mj}, super B factory~\cite{Hewett:2004tv},
super flavor factory~\cite{f-factory}, and LHCb~\cite{LHCb})
without caring about current constraints from $\meee$~\cite{Bellgardt:1987du}
and $\llg$~\cite{Brooks:1999pu, tlg}.
 It is a good feature of the A4HTM
that the model will be excluded
if $\meg$ is observed in ongoing MEG experiment~\cite{Cattaneo:2009qe}.
 Only $H_3^{\pm\pm}$ and $H_4^{\pm\pm}$ can give
a sizable $\tmee$ in this model
while $\temm$
(which is possible with $H_2^{\pm\pm}$)
can be affected also by neutral components
of doublet fields~\cite{A4NR}.
 Since $H_1^{\pm\pm}$ does not contribute to $\tlll$ also,
constraints on its coupling comes only from processes
given by $H_1^\pm$
if other $H_i^{\pm\pm}$ are heavy enough.
 The Bhabha scattering ($\bar{e} e \to \bar{e} e$)
can be contributed by $H_3^{\pm\pm}$ and $H_4^{\pm\pm}$.

\subsection{ $H_T^- \to \ell \nu$ }

%%%%%%%%%  Table for H+  %%%%%%%%%%%%
\begin{table}[t]
\begin{tabular}{c||c|c|c|c}
 {}
 & \ $\BR(H^-_T \to \ell\nu)$ \
 & \ $\mu \to e \bar{\nu}_\ell \nu_{\ell^\prime}$ \
 & \ $\tau \to \ell \bar{\nu}_\ell \nu_\tau$ \
 & \ matter effect, \
\\
 {}
 & \ $e\nu:\mu\nu:\tau\nu$
 & 
 & (coherent)
 & $\nu e \to \nu e$
\\\hline\hline
 $H^\pm_{T1}$
 & \ $1:1:4$
 & $\mu \to e \bar{\nu}_e \nu_\mu$
 & none
 & $\epsilon_{\mu\mu}^e$
\\\hline
 $H^\pm_{T2}$
 & \ $1:4:1$
 & $\mu \to e \bar{\nu}_\mu \nu_\tau$
 & $\tau \to e \bar{\nu}_e \nu_\tau$
 & $\epsilon_{\tau\tau}^e$
\\\hline
 $H^\pm_{T3}$
 & \ $2R^\pm_3 : 1 : 1$ \hspace*{4mm}
 & $\mu \to e \bar{\nu}_\tau \nu_e$
 & $\tau \to \mu \bar{\nu}_\mu \nu_\tau$
 & $\epsilon_{ee}^e$
\\\hline
 $H^\pm_{T4}$
 & \ $2R^\pm_4 : 1 : 1$ \hspace*{4mm}
 & $\mu \to e \bar{\nu}_\tau \nu_e$
 & $\tau \to \mu \bar{\nu}_\mu \nu_\tau$
 & $\epsilon_{ee}^e$
\end{tabular}
\caption{
 Ratios of decays of the triplet-like $H_{Ti}^\pm$
into a charged lepton and a neutrino
are summarized, where the flavors of neutrinos are summed up.
 Possible decays of $\mu$ with $H_{Ti}^\pm$ mediation
are also presented.
 The fourth column shows $\tau$ decays
which are coherent with the ones in the SM\@.
 The last column shows
contributions of $H_{Ti}^\pm$ to effective interactions
which relate to the non-standard matter effect
for the neutrino oscillation
and the elastic scattering of $\nu$
on the electron.
}
\label{tab:Hp}
\end{table}

%%%%%%%%  Phenomenology of H+ %%%%%%%%%%%%%%
 Table~\ref{tab:Hp} shows the processes
to which the triplet-like $H_{Ti}^\pm$ can contribute.
 The second column presents ratios of
the branching ratios $\BR_{\ell\nu} \equiv \BR(H^-_T \to \ell \nu)$
where the flavors of neutrinos in the final state are summed up.
For decays of $H^\pm_{T3}$ and $H^\pm_{T4}$
we used
\begin{eqnarray}
R^\pm_3
\equiv
 \frac{ |2h_\Delta c_\pm + \sqrt{3}\, \tilde{h}_\delta s_\pm|^2 }
      { 2 |h_\Delta c_\pm - \sqrt{3}\, \tilde{h}_\delta s_\pm|^2 } ,
\quad
%\\
%
%
%
R^\pm_4
\equiv
 \frac{ |2h_\Delta s_\pm - \sqrt{3}\, \tilde{h}_\delta c_\pm|^2 }
      { 2 |h_\Delta s_\pm + \sqrt{3}\, \tilde{h}_\delta c_\pm|^2 } ,
\end{eqnarray}
where $c_\pm \equiv \cos\theta_\pm$
and $s_\pm \equiv \sin\theta_\pm$.
 Similarly to the case for $H^{\pm\pm}$ decays,
$H^\pm_{T1}$ and $H^\pm_{T2}$ give
$\BR_{\mu\nu} \neq \BR_{\tau\nu}$
while the HTM gives $\BR_{\mu\nu} \simeq \BR_{\tau\nu}$.
 Decays of degenerate $H^\pm_{T1}$ and $H^\pm_{T2}$
result in $\BR_{\mu\nu} = \BR_{\tau\nu}$.
 It is found that
$\BR_{e\nu}$ can be larger than $\BR_{\mu\nu}$ ($=\BR_{\tau\nu}$)
for $H^\pm_{T3}$ and $H^\pm_{T4}$
although the neutrino masses $m_i$ in this model
give $\Delta m^2_{31} \equiv m_3^2 - m_1^2 > 0$.
 This is in contrast with
$\BR_{e\nu} < \BR_{\mu\nu}$ in the HTM for $\Delta m^2_{31} > 0$%
~\cite{Perez:2008ha}.

\subsection{$\mu \to e\bar{\nu}\nu$ and $\tau \to \ell \bar{\nu}\nu$}
 The third column of Table~\ref{tab:Hp} shows
$\mu \to e \bar{\nu}_\ell \nu_{\ell^\prime}$
which are possible with the $H^\pm_{Ti}$ mediation.
 It is important to note that $H^\pm_{Ti}$ can not contribute
to $\llg$ at one loop level.
 Only $H^\pm_{T1}$ gives the coherent decay
with the standard one of the $W$ boson exchange,
which can be larger effect than incoherent ones in principle.
 Of course,
we can not find anything new in the standard $\mu$ decay itself
because new effects are absorbed by the experimental definition
of the value of $G_F$.
 Incoherent ones given by other $H^\pm_{Ti}$ affect
measurements in the future neutrino factory
where the neutrino beam is produced by the $\mu$ decay.
 Neutrinos from the standard $\mu^-$ decay give signals of
$\mu^-$ and $e^+$ at the near detector.
 Non-standard effects on the neutrino production%
~\cite{Grossman:1995wx}
will be observed at the near detector
as the signals of the wrong-signed muon (for $H^\pm_{T2}$)
or the wrong-signed electron (for $H^\pm_{T3}$ and $H^\pm_{T4}$)
if the detector can discriminate the charge and flavors.

 The fourth column of Table~\ref{tab:Hp}
is for $\tau \to \ell \bar{\nu}_\ell \nu_\tau$
which are coherent with the decays via $W^\pm$ mediation.
 Note that each $H^\pm_{Ti}$ contribute
to a decay of $\mu$ or $\tau$
coherently with the $W^\pm$ contribution.
 Thus,
there can be a sizable difference between
effective couplings $G_{\mu e}$ ($\equiv G_F$) and $G_{\tau\ell}$
which are determined by $\mu \to e\bar{\nu}\nu$
and $\tau \to \ell \bar{\nu}\nu$, respectively.
 The effective coupling
$G_{\mu e}^2 \equiv \sum G_{\mu e\ell\ell^\prime}^2$
is given by the effective interactions
\begin{eqnarray}
2\sqrt{2}\, G_{\mu e\ell\ell^\prime}
 \left( \bar{\nu}_\ell \gamma_\mu P_L \mu \right)
 \left( \bar{e} \gamma^\mu P_L \nu_{\ell^\prime} \right) , 
\end{eqnarray}
and $G_{\tau\ell}$ are defined by the similar way.
 The contribution of $W^\pm$ to $G_{\mu e\ell\ell^\prime}$
is $G_{\mu e\ell\ell^\prime}^W \equiv g^2/(4\sqrt{2}\,m_W^2)$,
where $g$ denotes the gauge coupling constant of $SU(2)_L$
and $m_W$ is the mass of $W^\pm$.
 In the A4HTM,
contributions of $H_{Ti}^\pm$ to $G_{\mu e\ell\ell^\prime}$
can be expressed%
\footnote{
Note that
$2 (\overline{\nu_{\ell^\prime}^c} P_L \mu) (\overline{e} P_R \nu_\ell^c)
= (\overline{\nu}_\ell \gamma^\mu P_L \mu)
  (\overline{e} \gamma_\mu P_L \nu_{\ell^\prime})$.
}
as
\begin{eqnarray}
G_{\mu e\ell\ell^\prime}^{H_T^\pm}
\equiv
 \sum_i
 \frac{ (h_{i\pm})_{\ell^\prime \mu} (h_{i\pm}^\ast)_{\ell e}  }
      { 2\sqrt{2}\, m_{H_{Ti}^\pm}^2 }.
\end{eqnarray}

\subsection{ Non-standard interactions of neutrinos }

 During the propagation of neutrinos in the ordinary matter,
the coherent forward scattering of them
on the matter ($e$, $u$, and $d$)
affects neutrino oscillations%
~\cite{Wolfenstein:1977ue,Mikheev:1986gs}.
 The so-called non-standard interaction (NSI)
of neutrinos can give the non-standard matter effect
on the neutrino oscillation%
~\cite{Wolfenstein:1977ue,NSI-matter}.
 The relevant effective interaction for that is
\begin{eqnarray}
 2 \sqrt{2}\, G_F \epsilon_{\ell\ell^\prime}^{fP}
 \left( \overline{f} \gamma^\mu P f \right)
 \left( \bar{\nu}_\ell \gamma_\mu P_L \nu_{\ell^\prime} \right),
\label{eq:NSI}
\end{eqnarray}
where $f=e, u ,d$ and $P= P_L, P_R$.
 The interaction eq.~(\ref{eq:NSI}) is defined just for the non-standard one,
which should be added to the standard one of the weak interaction.
 Although eq.~(\ref{eq:NSI}) is written
in the form of the neutral current interaction,
the effective interaction can be given by
the charged scalar mediation also
because of the Fierz transformation%
\footnote{
Note that
$2 (\overline{\nu_{\ell^\prime}^c} P_L e) (\overline{e} P_R \nu_\ell^c)
= (\overline{e} \gamma^\mu P_L e)
   (\overline{\nu_\ell} \gamma_\mu P_L \nu_{\ell^\prime})$.
}.
 The triplet-like $H_{Ti}^\pm$ in the A4HTM
can generate
$\epsilon_{\ell\ell^\prime}^{fP}$
with only the left-handed electron
for only $\ell=\ell^\prime$ as
\begin{eqnarray}
\epsilon_{\ell\ell}^{eP_L}
=
 \sum_i
 \frac{ | (h_{i\pm})_{e\ell} |^2 }
      { 2\sqrt{2} G_F m_{H_{Ti}^\pm}^2 } .
\end{eqnarray}
 The last column of Table~\ref{tab:Hp} shows
$\epsilon_{\ell\ell}^{eP_L}$ induced by each $H_{Ti}^\pm$.
 Possible sizes of $\epsilon_{\ell\ell}^{eP_L}$
are shown in the next section
by considering other constraints.
 Contributions of the doublet like charged Higgs fields
to $\epsilon_{\ell\ell}^{eP_R}$ are negligible
because Yukawa couplings appear as $m_e^2/v^2$.
 The elastic scattering of neutrinos on the electron
is affected also by $\epsilon_{\ell\ell^\prime}^{eP_L}$.
 A study on the NSI in the HTM for the matter effect
and the neutrino production (See the previous subsection also)
can be seen in \cite{Malinsky:2008qn}.
 Model-independent constraints on the NSI for the matter effect
can be found in \cite{Davidson:2003ha}.

\subsection{ Doublet Higgs sector }
 Contributions of doublet-like Higgs bosons
to the flavor violating decays of charged leptons
are the same as the ones in a model discussed
in \cite{A4NR}
(See also \cite{Ma:2008ym}).
 Two combinations
$\Phi_\eta
 \equiv (\Phi_x + \omega^2 \Phi_y + \omega \Phi_z)/\sqrt{3}$
and
$\Phi_\zeta
 \equiv (\Phi_x + \omega \Phi_y + \omega^2 \Phi_z)/\sqrt{3}$,
which have no vev and no contribution
to the mass matrix of charged leptons,
can cause flavor changing neutral currents.
 The largest contribution of
doublet-like neutral Higgs bosons
(real and imaginary parts of
$(\phi_\eta^0+\phi_\zeta^0)/\sqrt{2}$
and $(-i\phi_\eta^0+i\phi_\zeta^0)/\sqrt{2}$)
is to $\tau_R \to \overline{e_L} \mu_L \mu_R$
for which the Yukawa coupling appears
as $m_\mu m_\tau/v^2$.
 There is no contribution to $\meee$
and $\ell \to \ell^\prime \gamma$
because of an approximate $Z_3$ symmetry.
 The quark sector can be just like the SM one,
which is described by only {\bf 1}-representations
with an additional Higgs doublet field $\Phi_q$
as mentioned in \cite{A4NR}.
 The phenomenology of $\Phi_q$
and $\Phi_\xi \equiv (\Phi_x + \Phi_y + \Phi_z)/\sqrt{3}$
is almost identical to a type of the two-Higgs-doublet-models,
which can be seen in
\cite{Barger:1989fj, Grossman:1994jb, Akeroyd:1994ga, Aoki:2009ha}.

%%%%%%%%%%%%%%   sect: constraints  %%%%%%%%%%%
\section{Constraints}\label{sect:constraints}

 In this section,
constraints on the model
and future prospects are considered.
 We assume that
one of $H_i^{\pm\pm}$ is much lighter than the others
for simplicity.
 Then,
one of $H_{Ti}^\pm$ should be light also
because large mass splittings are disfavored
by the $\rho$ parameter.

\subsection{Case of light $H_1^{\pm\pm}$ and $H_{T1}^\pm$}
 If only $H_1^{\pm\pm}$ is light enough among $H_i^{\pm\pm}$,
there is no constraint on the model from $\tlll$.
 Since $H_{T1}^\pm$ also must be light enough in this case,
a constraint comes from
\begin{eqnarray}
\frac{G_{\tau e}^2}{G_F^2}
=
 \frac{ (G^W)^2 }{ ( G^W + G^{H_T^\pm}_{\mu e} )^2 }
= \frac{ (G_F - G^{H_T^\pm}_{\mu e})^2 }{ G_F^2 }
= 1.0012 \pm 0.0053 \ \
(\text{p.~512 of \cite{Amsler:2008zzb}}),
\label{eq:GF-H1}
\end{eqnarray}
where $G^W$ and $G^{H_T^\pm}_{\ell\ell^\prime}$ indicate
contributions of $W$ and $H_T^\pm$ to $G_{\ell\ell^\prime}$,
respectively.
 We obtain
\begin{eqnarray}
|h_\Delta|^2
<
 3.4\times 10^{-2}
 \left( \frac{ m_{H_{T1}^\pm} }{ 300\,\GeV } \right)^2
\ \ \text{(90\%CL)}
\label{eq:hD-H1}
\end{eqnarray}

 The coefficient of NSI
relevant to the matter effect for the neutrino oscillation 
is constrained by eq.~(\ref{eq:hD-H1}) as
$|\epsilon_{\mu\mu}^e| = |G^{H_T^\pm}_{\mu e}/G_F|
 < 3.8\times 10^{-3}$
which is smaller than
the expected sensitivity ($\sim 0.1$)~\cite{NSI-nufact}
in the neutrino factory.
 There is no effect on the production of the neutrino beam.

\subsection{Case of light $H_2^{\pm\pm}$ and $H_{T2}^\pm$}
 If $H_2^{\pm\pm}$ is lighter enough than other $H_i^{\pm\pm}$,
a constraint on the model is given by
\begin{eqnarray}
\BR(\temm)
= \frac{ |h_\Delta|^4 }{ 36 G_F^2 m_{H_{T2}^{\pm\pm}}^4 }
  \BR( \tau \to \mu \bar{\nu}_\mu \nu_\tau)
< 1.7\times 10^{-8} \ \ \text{(90\%CL)~\cite{Hayasaka:2010np}} ,
\end{eqnarray}
where $\BR( \tau \to \mu \bar{\nu}_\mu \nu_\tau) \simeq 0.17$.
 We have
\begin{eqnarray}
| h_\Delta |^2
< 2.0\times 10^{-3}
  \left( \frac{ m_{H_2^{\pm\pm}} }{ 300\,\GeV } \right)^2 \ \
\text{(90\%CL)} .
\label{eq:hD-H2}
\end{eqnarray}
 Another constraint on $|h_\Delta|$ can be obtained
by $G_{\tau e}^2/G_F^2 = 1.0012 \pm 0.0053$ as
\begin{eqnarray}
|h_\Delta|^2
< 4.4\times 10^{-2}
  \left( \frac{ m_{H_{T2}^\pm} }{ 300\,\GeV } \right)^2 \ \
\text{(90\%CL)} ,
\end{eqnarray}
although this is weaker than eq.~(\ref{eq:hD-H2})
because $m_{H_{T2}^\pm}$ should not be
very different from $m_{H_2^{\pm\pm}}$.

 The effective coupling $G_{\mu e e \tau}$
for $\mu \to e \bar{\nu}_e \nu_\tau$
is constrained by eq.~(\ref{eq:hD-H2})
with $m_{H_{T2}^\pm} \simeq m_{H_2^{\pm\pm}}$
as $|G_{\mu e e \tau}/G_F| \lesssim 2\times 10^{-4}$
which can be around the expected sensitivity
at a near detector of the neutrino factory%
~\cite{Malinsky:2008qn}.
 The non-standard matter effect with $\epsilon_{\tau\tau}^e$
is too small to be observed in the neutrino factory
because eq.~(\ref{eq:hD-H2}) results in
$\epsilon_{\tau\tau}^e \lesssim 10^{-3}$.

\subsection{Case of light $H_3^{\pm\pm}$ and $H_{T3}^\pm$}
 Let us remind that we have defined as
$m_{H_3^{\pm\pm}} \leq m_{H_4^{\pm\pm}}$.
 If $H_4^{\pm\pm}$ is very heavy,
a relevant constraint is
\begin{eqnarray}
\BR(\tmee)
= \frac{
        \bigl|
         (h_{3\pm\pm})_{\tau\mu}\,
         (h_{3\pm\pm})_{ee}
        \bigr|^2
       }
       { 4 G_F^2 m_{H_3^{\pm\pm}}^4 }
  \BR( \tau \to \mu \bar{\nu}_\mu \nu_\tau)
< 1.5\times 10^{-8}
\ \ \text{(90\%CL)~\cite{Hayasaka:2010np}} ,
\end{eqnarray}
which results in
\begin{eqnarray}
\bigl|(h_{3\pm\pm})_{\tau\mu}\, (h_{3\pm\pm})_{ee}\bigr|
< 6.3\times 10^{-4}
  \left( \frac{ m_{H_3^{\pm\pm}} }{ 300\,\GeV } \right)^2
\ \ \text{(90\%CL)} .
\label{eq:h3tmee}
\end{eqnarray}
 The constraint on $|(h_{3\pm\pm})_{ee}|$ itself
is given by the Bhabha scattering~\cite{Bhabha}.
 For example,
we have%
\footnote
{
 The bound at 95\%CL in \cite{Bhabha}
is translated naively to the bound at 90\%CL
by a factor of $1.9/1.6$,
where 95\%CL and 90\%CL
correspond to $1.9\sigma$ and $1.6\sigma$,
respectively.
}
\begin{eqnarray}
|(h_{3\pm\pm})_{ee}|\lesssim 0.3
\ \ \text{(90\%CL, $m_{H_3^{\pm\pm}}=300\,\GeV$)} .
\label{eq:h3ee}
\end{eqnarray}
 For $h_{3\pm}$,
a constraint comes from
$G_{\tau\mu}^2/G_F^2=0.981 \pm 0.018$
(p.~512 of \cite{Amsler:2008zzb}),
and we have
\begin{eqnarray}
\bigl|(h_{3\pm})_{\tau\mu}\bigr|^2
< 1.6\times 10^{-3}
  \left( \frac{ m_{H_{T3}^\pm} }{ 300\,\GeV } \right)^2
\ \ \text{(90\%CL)} .
\end{eqnarray}
 The LSND result~\cite{Auerbach:2001wg}
on $\nu_e e$ elastic scattering,
$\sigma_{\nu_e e}^{\text{LSND}} =
(10.1 \pm 1.5) E_{\nu_e}(\MeV)\times 10^{-45} \text{cm}^2$,
can be translated into a constraint
$\epsilon_{ee}^{eP_L} < 0.11$ at 90\%CL~\cite{Davidson:2003ha}.
 A comparable constraint on $\epsilon_{ee}^{eP_L}$
was obtained with solar and reactor neutrinos~\cite{Bolanos:2008km}.
 The constraint $\epsilon_{ee}^{eP_L} < 0.11$ can be written as
\begin{eqnarray}
\bigl|(h_{3\pm})_{ee}\bigr|^2
< 0.33 \left( \frac{ m_{H_{T3}^\pm} }{ 300\,\GeV } \right)^2
\ \ \text{(90\%CL)} .
\label{eq:h3ee-2}
\end{eqnarray}

 For $m_{H_{T3}^\pm} \simeq m_{H_3^{\pm\pm}}$
(namely, $|\lambda_5| \ll 1$ and then $h_{i\pm}\simeq h_{i\pm\pm}$),
the effective coupling $G_{\mu e \tau e}$ for
$\mu \to e \bar{\nu}_\tau \nu_e$
is constrained by eq.~(\ref{eq:h3tmee}) as
$G_{\mu e \tau e}/G_F \lesssim 2\times 10^{-4}$.
 Constraints of eq.~(\ref{eq:h3ee}) and (\ref{eq:h3ee-2})
on the non-standard matter effect are comparable
($\epsilon_{ee}^{eP_L} \lesssim 0.1$).
 These non-standard effects
can be close to the expected sensitivity
in the neutrino factory.

%%%%%%%%%%   sect: conclusion  %%%%%%%%%%
\section{Conclusions}\label{sect:concl}
 In this article,
we investigated the phenomenology
of triplet Higgs bosons
in the simplest $A_4$-symmetric version
of the Higgs Triplet Model (A4HTM).
 The A4HTM is a four-Higgs-Triplet-Model
($\delta$ of ${\bf 1}$ and
($\Delta_x$, $\Delta_y$, $\Delta_z$) of ${\bf 3}$).
 Four mass eigenstates of
doubly charged Higgs bosons, $H_i^{\pm\pm}$,
are obtained explicitly
from the Higgs potential.
 We also obtained four mass eigenstates
of the triplet-like singly charged Higgs bosons, $H_{Ti}^\pm$,
for which doublet components can be ignored
because of small triplet vev's.

 It was shown that
the A4HTM gives unique predictions about
their decay branching ratios into two leptons
($H_i^{--} \to \ell \ell^\prime$ and $H_{iT}^- \to \ell \nu$);
 for example,
the leptonic decays of $H_2^{--}$
are only into $\mu\mu$ and $e\tau$
because an approximate $Z_3$ symmetry remains,
and the ratio of the branching ratios is \text{$2:1$}
as a consequence of the $A_4$ symmetry
in the original Lagrangian.
 Therefore,
it will be possible to test the model
at hadron colliders (Tevatron and LHC)
if some of these Higgs bosons are light enough
to be produced.

 Even if these Higgs bosons are too heavy
to be produced at hadron colliders,
they can affect the lepton flavor violating decays
of charged leptons
if the triplet Yukawa coupling constants are large enough.
 It was shown that
there is no contribution of these Higgs bosons
to $\meee$ and $\llg$.
 Thus,
we can naturally expect signals of $\tmee$ and $\temm$
(which are possible in this model among six $\tlll$)
in the future in collider experiments
(Super-KEKB, super B factory, super flavor factory, and LHCb)
without interfering with a stringent experimental bound
on $\meee$.
 This model will be excluded
if $\llg$ is observed.

 We considered current experimental constraints on the model
and prospects of the measurement
of the non-standard neutrino interactions (NSI)
in the neutrino factory.
 If $H_2^{\pm\pm}$ or $H_3^{\pm\pm}$
is lighter enough than other $H_i^{\pm\pm}$,
effects of the NSI can be around
the expected sensitivity in the neutrino factory.

\begin{acknowledgments}
The work of T.F.\ is supported in part by the Grant-in-Aid for Science Research
from the Ministry of Education, Science and Culture of Japan
(No.~020540282 and No.~21104004).
\end{acknowledgments}

%%%%%%%%%  appendix  %%%%%%%%%%%%%%%
\appendix

%%%%%  decomposition  %%%%%%%%%%
\section{Decompositions}\label{app:A4}

 For $a = (a_x, a_y, a_z)^T$ and $b = (b_x, b_y, b_z)^T$
of ${\bf 3}$ in the $S$-diagonal basis of eq.~(\ref{eq:sdiag}),
we used
\begin{eqnarray}
(a b)_{\bf 1}
&\equiv&
 a_x b_x + a_y b_y + a_z b_z ,\\
(a b)_{{\bf 1}^\prime}
&\equiv&
 a^T X^\prime b
=
 a_x b_x + \omega^2 a_y b_y + \omega a_z b_z , \quad
X^\prime
\equiv
 \begin{pmatrix}
  1 & 0 & 0\\
  0 & \omega^2 & 0\\
  0 & 0 & \omega
 \end{pmatrix} ,
\\
(a b)_{{\bf 1}^\dprime}
&\equiv&
 a^T X^\dprime b
=
 a_x b_x + \omega a_y b_y + \omega^2 a_z b_z , \quad
X^\dprime
\equiv
 \begin{pmatrix}
  1 & 0 & 0\\
  0 & \omega & 0\\
  0 & 0 & \omega^2
 \end{pmatrix} ,
\\
(a b)_{{\bf 3}_s}
&\equiv&
 \begin{pmatrix}
  a^T V_{sx} b , \
  a^T V_{sy} b , \
  a^T V_{sz} b
 \end{pmatrix}^T
\nonumber\\
&=&
 \begin{pmatrix}
  a_y b_z + a_z b_y , \
  a_z b_x + a_x b_z , \
  a_x b_y + a_y b_x
 \end{pmatrix}^T ,\\
&&
V_{sx}
\equiv
 \begin{pmatrix}
  0 & 0 & 0\\
  0 & 0 & 1\\
  0 & 1 & 0
 \end{pmatrix} , \quad
V_{sy}
\equiv
 \begin{pmatrix}
  0 & 0 & 1\\
  0 & 0 & 0\\
  1 & 0 & 0
 \end{pmatrix} , \quad
V_{sz}
\equiv
 \begin{pmatrix}
  0 & 1 & 0\\
  1 & 0 & 0\\
  0 & 0 & 0
 \end{pmatrix} ,
\\
(a b)_{{\bf 3}_a}
&\equiv&
 \begin{pmatrix}
  a^T V_{ax} b , \
  a^T V_{ay} b , \
  a^T V_{az} b
 \end{pmatrix}^T
\nonumber\\
&=&
 \begin{pmatrix}
  a_y b_z - a_z b_y , \
  a_z b_x - a_x b_z , \
  a_x b_y - a_y b_x
 \end{pmatrix}^T ,\\
&&
V_{ax}
\equiv
 \begin{pmatrix}
  0 & 0 & 0\\
  0 & 0 & 1\\
  0 & -1 & 0
 \end{pmatrix} , \quad
V_{ay}
\equiv
 \begin{pmatrix}
  0 & 0 & -1\\
  0 & 0 & 0\\
  1 & 0 & 0
 \end{pmatrix} , \quad
V_{az}
\equiv
 \begin{pmatrix}
  0 & 1 & 0\\
  -1 & 0 & 0\\
  0 & 0 & 0
 \end{pmatrix} .
\end{eqnarray}

 If we use a $T$-diagonal basis defined as
\begin{eqnarray}
{\bf 3}_T
&\equiv&
 U_T {\bf 3}, \quad
U_T
\equiv
 \frac{1}{\sqrt{3}}
 \begin{pmatrix}
  1 &        1 & 1\\
  1 & \omega^2 & \omega\\
  1 &   \omega & \omega^2
 \end{pmatrix} ,\\
\tilde{S}\, {\bf 3}_T
&=&
   \frac{1}{\,3\,}
   \begin{pmatrix}
    -1 &  2 &  2\\
     2 & -1 &  2\\
     2 &  2 & -1
   \end{pmatrix}
   {\bf 3}_T, \quad
\tilde{T}\, {\bf 3}_T
 = \begin{pmatrix}
    1 & 0 & 0\\
    0 & \omega & 0\\
    0 & 0 & \omega^2
   \end{pmatrix}
   {\bf 3}_T ,
\label{eq:tdiag}
\end{eqnarray}
there are two kinds of decompositions:
${\bf 3}_T \otimes {\bf 3}_T
= {\bf 1} \oplus {\bf 1}^\prime \oplus {\bf 1}^\dprime
  \oplus {\bf 3}_{Ts} \oplus {\bf 3}_{Ta}$
and
${\bf 3}_T^\ast \otimes {\bf 3}_T
= {\bf 1} \oplus {\bf 1}^\prime \oplus {\bf 1}^\dprime
  \oplus {\bf 3}_T \oplus {\bf 3}_T^\ast$.
 Note that
${\bf 3}_T^\ast \otimes {\bf 3}_T^\ast
= ( {\bf 3}_T \otimes {\bf 3}_T )^\ast$.
%
%
%%%%%%%%%%% 3 * 3 in T-diagonal %%%%%%%%%%
 For $a_T \equiv (a_\xi, a_\eta, a_\zeta)^T$
and $b_T \equiv (b_\xi, b_\eta, b_\zeta)^T$
in the $T$-diagonal basis,
decompositions for ${\bf 3}_T \otimes {\bf 3}_T$ are given by
\begin{eqnarray}
{\bf 3}_T \otimes {\bf 3}_T \to {\bf 1}
&:&
 a_T^T\, \Xi_s\, b_T
=
 a_\xi b_\xi + a_\eta b_\zeta + a_\zeta b_\eta , \quad
\Xi_s
\equiv
 \begin{pmatrix}
  1 & 0 & 0\\
  0 & 0 & 1\\
  0 & 1 & 0
 \end{pmatrix} ,
\\
{\bf 3}_T \otimes {\bf 3}_T \to {\bf 1}^\prime
&:&
 a_T^T\, \Xi_s^\prime\, b_T
=
 a_\xi b_\eta + a_\eta b_\xi + a_\zeta b_\zeta , \quad
\Xi_s^\prime
\equiv
 \begin{pmatrix}
  0 & 1 & 0\\
  1 & 0 & 0\\
  0 & 0 & 1
 \end{pmatrix} ,
\\
{\bf 3}_T \otimes {\bf 3}_T \to {\bf 1}^\dprime
&:&
 a_T^T\, \Xi_s^\dprime\, b_T
=
 a_\xi b_\zeta + a_\eta b_\eta + a_\zeta b_\xi , \quad
\Xi_s^\dprime
\equiv
 \begin{pmatrix}
  0 & 0 & 1\\
  0 & 1 & 0\\
  1 & 0 & 0
 \end{pmatrix} ,
\\
{\bf 3}_T \otimes {\bf 3}_T \to {\bf 3}_{Ts}
&:&
 \begin{pmatrix}
  a_T^T\, V_{s\xi}\, b_T , \
  a_T^T\, V_{s\eta}\, b_T , \
  a_T^T\, V_{s\zeta}\, b_T
 \end{pmatrix}^T ,\\
V_{s\xi}
&\equiv&
 \begin{pmatrix}
  2 &  0 & 0\\
  0 &  0 & -1\\
  0 & -1 & 0
 \end{pmatrix} , \
V_{s\eta}
 \equiv
 \begin{pmatrix}
   0 & -1 & 0\\
  -1 &  0 & 0\\
   0 &  0 & 2
 \end{pmatrix} , \
V_{s\zeta}
 \equiv
 \begin{pmatrix}
   0 & 0 & -1\\
   0 & 2 & 0\\
  -1 & 0 & 0
 \end{pmatrix} ,\\
{\bf 3}_T \otimes {\bf 3}_T \to {\bf 3}_{Ta}
&:&
 \begin{pmatrix}
  a_T^T\, V_{a\xi}\, b_T , \
  a_T^T\, V_{a\eta}\, b_T , \
  a_T^T\, V_{a\zeta}\, b_T
 \end{pmatrix}^T ,\\
V_{a\xi}
&\equiv&
 \begin{pmatrix}
  0 & 0 & 0\\
  0 & 0 & -i\\
  0 & i & 0
 \end{pmatrix} , \
V_{a\eta}
 \equiv
 \begin{pmatrix}
  0 & -i & 0\\
  i &  0 & 0\\
  0 &  0 & 0
 \end{pmatrix} , \
V_{a\zeta}
 \equiv
 \begin{pmatrix}
   0 & 0 & i\\
   0 & 0 & 0\\
  -i & 0 & 0
 \end{pmatrix} .
\end{eqnarray}
%
%
%
%%%%%%%%%%%%  3^dagger * 3  in T-diagonal %%%%%%%%%%%%%%
 On the other hand,
decompositions for ${\bf 3}_T^\ast \otimes {\bf 3}_T$ are given by
\begin{eqnarray}
{\bf 3}_T^\ast \otimes {\bf 3}_T \to {\bf 1}
&:&
 a_T^\dagger
 \begin{pmatrix}
  1 & 0 & 0\\
  0 & 1 & 0\\
  0 & 0 & 1
 \end{pmatrix}
 b_T
=
 a_\xi^\ast b_\xi + a_\eta^\ast b_\eta + a_\zeta^\ast b_\zeta ,\\
{\bf 3}_T^\ast \otimes {\bf 3}_T \to {\bf 1}^\prime
&:&
 a_T^\dagger\, \Xi^\prime\, b_T
=
 a_\xi^\ast b_\eta + a_\eta^\ast b_\zeta + a_\zeta^\ast b_\xi , \quad
\Xi^\prime
\equiv
 \begin{pmatrix}
  0 & 1 & 0\\
  0 & 0 & 1\\
  1 & 0 & 0
 \end{pmatrix} ,\\
{\bf 3}_T^\ast \otimes {\bf 3}_T \to {\bf 1}^\dprime
&:&
 a_T^\dagger\, \Xi^\dprime\, b_T
=
 a_\xi^\ast b_\zeta + a_\eta^\ast b_\xi + a_\zeta^\ast b_\eta , \quad
\Xi^\dprime
\equiv
 \begin{pmatrix}
  0 & 0 & 1\\
  1 & 0 & 0\\
  0 & 1 & 0
 \end{pmatrix} ,\\
{\bf 3}_T^\ast \otimes {\bf 3}_T \to {\bf 3}_T
&:&
 \begin{pmatrix}
  a_T^\dagger\, V_\xi\, b_T , \
  a_T^\dagger\, V_\eta\, b_T , \
  a_T^\dagger\, V_\zeta\, b_T
 \end{pmatrix}^T ,\\
{\bf 3}_T^\ast \otimes {\bf 3}_T \to {\bf 3}_T^\ast
&:&
 \begin{pmatrix}
  a_T^\dagger\, V_\xi^\ast\, b_T , \
  a_T^\dagger\, V_\eta^\ast\, b_T , \
  a_T^\dagger\, V_\zeta^\ast\, b_T
 \end{pmatrix}^T ,\\
V_\xi
&\equiv&
 \begin{pmatrix}
  1 & 0 & 0\\
  0 & \omega & 0\\
  0 & 0 & \omega^2
 \end{pmatrix} , \
V_\eta
 \equiv
 \begin{pmatrix}
       0 & \omega^2 & 0\\
       0 &        0 & 1\\
  \omega &      0 & 0
 \end{pmatrix} , \
V_\zeta
 \equiv
 \begin{pmatrix}
  0 & 0 & \omega\\
  \omega^2 & 0 & 0\\
  0 & 1 & 0
 \end{pmatrix} .
\end{eqnarray}

%%%%%%%  Fierz transformation  %%%%%%%%%
\section{"Fierz transformation"}

 We show useful relations to construct
the $A_4$-symmetric Higgs potential,
which are similar to the famous Fierz transformation
for the four-fermion interactions.
 We need not to use the relations explicitly
but we should keep the existence in our mind
in order to reduce the number of terms in the Higgs potential.
 Let $\phi_i$ ($i=1\,\text{--}\,4$) wave functions of ${\bf 3}$
in the basis of eq.~(\ref{eq:sdiag}).
 For terms involving three $\phi_i$,
we have
\begin{eqnarray}
&&
\begin{pmatrix}
 \bigl[ \phi_1 ( \phi_2 \phi_3 )_{{\bf 3}_s} \bigr]_{\bf 1} \\
 \bigl[ \phi_1 ( \phi_2 \phi_3 )_{{\bf 3}_a} \bigr]_{\bf 1}
\end{pmatrix}
=
 \begin{pmatrix}
  1 & 0\\
  0 & -1
 \end{pmatrix}
 \begin{pmatrix}
  \bigl[ ( \phi_1 \phi_2 )_{{\bf 3}_s} \phi_3 \bigr]_{\bf 1} \\
  \bigl[ ( \phi_1 \phi_2 )_{{\bf 3}_a} \phi_3 \bigr]_{\bf 1}
 \end{pmatrix} .
\end{eqnarray}
 Thus,
we can concentrate ourselves to one of the sets of the decompositions,
$\phi_1 (\phi_2 \phi_3)$ or $(\phi_1 \phi_2) \phi_3$.
 Similar relations for the term involving four $\phi_i$
are obtained as
\begin{eqnarray}
&&\hspace*{-10mm}
\bigg(
 ( \phi_1 \phi_2 )_{\bf 1}
  ( \phi_3 \phi_4 )_{\bf 1}, \
 ( \phi_1 \phi_2 )_{{\bf 1}^\prime}
  ( \phi_3 \phi_4 )_{{\bf 1}^\dprime}, \
 ( \phi_1 \phi_2 )_{{\bf 1}^\dprime}
  ( \phi_3 \phi_4 )_{{\bf 1}^\prime}, \
 \Bigl(
  ( \phi_1 \phi_2 )_{{\bf 3}_s} ( \phi_3 \phi_4 )_{{\bf 3}_s}
 \Bigr)_{\bf 1},
\nonumber\\
&&\hspace*{10mm}
 \Bigl(
  ( \phi_1 \phi_2 )_{{\bf 3}_a} ( \phi_3 \phi_4 )_{{\bf 3}_a}
 \Bigr)_{\bf 1}, \
 \Bigl(
  ( \phi_1 \phi_2 )_{{\bf 3}_s} ( \phi_3 \phi_4 )_{{\bf 3}_a}
 \Bigr)_{\bf 1}, \
 \Bigl(
  ( \phi_1 \phi_2 )_{{\bf 3}_a} ( \phi_3 \phi_4 )_{{\bf 3}_s}
 \Bigr)_{\bf 1}
\bigg)^T
%\begin{pmatrix}
% ( \phi_1 \phi_2 )_{\bf 1} ( \phi_3 \phi_4 )_{\bf 1} \\
% ( \phi_1 \phi_2 )_{{\bf 1}^\prime} ( \phi_3 \phi_4 )_{{\bf 1}^\dprime}\\
% ( \phi_1 \phi_2 )_{{\bf 1}^\dprime} ( \phi_3 \phi_4 )_{{\bf 1}^\prime}\\
% \Bigl(
%  ( \phi_1 \phi_2 )_{{\bf 3}_s} ( \phi_3 \phi_4 )_{{\bf 3}_s}
% \Bigr)_{\bf 1} \\
% \Bigl(
%  ( \phi_1 \phi_2 )_{{\bf 3}_a} ( \phi_3 \phi_4 )_{{\bf 3}_a}
% \Bigr)_{\bf 1} \\
% \Bigl(
%  ( \phi_1 \phi_2 )_{{\bf 3}_s} ( \phi_3 \phi_4 )_{{\bf 3}_a}
% \Bigr)_{\bf 1} \\
% \Bigl(
%  ( \phi_1 \phi_2 )_{{\bf 3}_a} ( \phi_3 \phi_4 )_{{\bf 3}_s}
% \Bigr)_{\bf 1}
%\end{pmatrix}
\nonumber\\
&&\hspace*{0mm}
=
 \frac{1}{12}
 \begin{pmatrix}
  4
   & 4
   & 4
   & 6
   & -6
   & 0
   & 0\\
  4
   & 4
   & 4
   & -3
   & 3
   & -3i\sqrt{3}
   & 3i\sqrt{3}\\
  4
   & 4
   & 4
   & -3
   & 3
   & 3i\sqrt{3}
   & -3i\sqrt{3}\\
  8
   & -4
   & -4
   & 6
   & 6
   & 0
   & 0\\
  -8
   & 4
   & 4
   & 6
   & 6
   & 0
   & 0\\
   0
   & 4 i \sqrt{3}
   & -4 i \sqrt{3}
   & 0
   & 0
   & 6
   & 6\\
   0
   & -4 i \sqrt{3}
   & 4 i \sqrt{3}
   & 0
   & 0
   & 6
   & 6
 \end{pmatrix}
 \begin{pmatrix}
  ( \phi_1 \phi_4 )_{\bf 1} ( \phi_3 \phi_2 )_{\bf 1}\\
  ( \phi_1 \phi_4 )_{{\bf 1}^\prime} ( \phi_3 \phi_2 )_{{\bf 1}^\dprime}\\
  ( \phi_1 \phi_4 )_{{\bf 1}^\dprime} ( \phi_3 \phi_2 )_{{\bf 1}^\prime}\\
  \Bigl(
   ( \phi_1 \phi_4 )_{{\bf 3}_s} ( \phi_3 \phi_2 )_{{\bf 3}_s}
  \Bigr)_{\bf 1}\\
  \Bigl(
   ( \phi_1 \phi_4 )_{{\bf 3}_a} ( \phi_3 \phi_2 )_{{\bf 3}_a}
  \Bigr)_{\bf 1}\\
  \Bigl(
   ( \phi_1 \phi_4 )_{{\bf 3}_s} ( \phi_3 \phi_2 )_{{\bf 3}_a}
  \Bigr)_{\bf 1}\\
  \Bigl(
   ( \phi_1 \phi_4 )_{{\bf 3}_a} ( \phi_3 \phi_2 )_{{\bf 3}_s}
  \Bigr)_{\bf 1}
 \end{pmatrix} .
\end{eqnarray}
 These relations are obtained by
the "Fierz transformation" for $3\times 3$ matrices:
\begin{eqnarray}
&&
( \phi_1 \Gamma^i \phi_2 )
( \phi_3 (\Gamma^j)^\dagger \phi_4 )
=
 \sum_k
 ( \phi_1 M_{ij}^k \phi_4 )
 ( \phi_3 (\Gamma^k)^\dagger \phi_2 ) , \quad
M_{ij}^k
\equiv
 \Gamma^i \Gamma^k (\Gamma^j)^\dagger ,
\\
&&
\Gamma^i
\equiv
\left\{
 \frac{1}{\sqrt{3}} I, \
 \frac{1}{\sqrt{3}} X^\prime, \
 \frac{1}{\sqrt{3}} X^\dprime, \
 \frac{1}{\sqrt{2}} V_{sx}, \
 \frac{1}{\sqrt{2}} V_{sy}, \
 \frac{1}{\sqrt{2}} V_{sz},
\right.
\nonumber\\
&&
\hspace*{70mm}
\left.
 \frac{1}{\sqrt{2}} V_{ax}, \
 \frac{1}{\sqrt{2}} V_{ay}, \
 \frac{1}{\sqrt{2}} V_{az}
\right\} ,
\end{eqnarray}
where $I$ is the identity matrix
and $\Gamma^i$ give the complete set
of $3\times 3$ matrices
which satisfy
$\tr(\Gamma^i (\Gamma^j)^\dagger) = \delta^{ij}$.
 Definitions of the matrices of $\Gamma^i$
are shown in Appendix~\ref{app:A4}.

%%%%%%%  Higgs potential  %%%%%%%%
\section{Higgs Potential}\label{app:V}

 We show for completeness
the parts of the $A_4$-symmetric Higgs potential,
which are not used in the main part of this article:
\begin{eqnarray}
&&
V_1
=
 \lambda_{1}
 \bigl[ (\Phi^\dagger \Phi)_{\bf 1} \bigr]^2
 +
 \lambda_{1p}
 (\Phi^\dagger \Phi)_{{\bf 1}^\prime}
 (\Phi^\dagger \Phi)_{{\bf 1}^\dprime}
\nonumber\\
&&\hspace*{10mm}
{}
 +
 \lambda_{1ss}
 \bigl(
  (\Phi^\dagger \Phi)_{{\bf 3}_s}
  (\Phi^\dagger \Phi)_{{\bf 3}_s}
 \bigr)_{\bf 1}
 +
 \lambda_{1aa}
 \bigl(
  (\Phi^\dagger \Phi)_{{\bf 3}_a}
  (\Phi^\dagger \Phi)_{{\bf 3}_a}
 \bigr)_{\bf 1}
\nonumber\\
&&\hspace*{10mm}
{}
 +
 i \lambda_{1sa}
 (\Phi^\dagger \Phi)_{{\bf 3}_s}
 (\Phi^\dagger \Phi)_{{\bf 3}_a} ,
\end{eqnarray}

\begin{eqnarray}
&&
V_2
=
 \lambda_{2\delta}
 \bigl[ \tr(\delta^\dagger \delta) \bigr]^2
\nonumber\\
&&\hspace*{10mm}
{}
 +
 \lambda_{2\Delta}
 \bigl[ \tr(\Delta^\dagger \Delta)_{\bf 1} \bigr]^2
 +
 \lambda_{2\Delta p}
 \tr(\Delta^\dagger \Delta)_{{\bf 1}^\prime}\,
 \tr(\Delta^\dagger \Delta)_{{\bf 1}^\dprime}
\nonumber\\
&&\hspace*{10mm}
{}
 +
 \lambda_{2\Delta ss}\,
 \Bigl(
  \tr(\Delta^\dagger \Delta)_{{\bf 3}_s}\,
  \tr(\Delta^\dagger \Delta)_{{\bf 3}_s}
 \Bigr)_{\bf 1}
 +
 \lambda_{2\Delta aa}\,
 \Bigl(
  \tr(\Delta^\dagger \Delta)_{{\bf 3}_a}\,
  \tr(\Delta^\dagger \Delta)_{{\bf 3}_a}
 \Bigr)_{\bf 1}
\nonumber\\
&&\hspace*{10mm}
{}
 +
 i \lambda_{2\Delta sa}\,
 \Bigl(
  \tr(\Delta^\dagger \Delta)_{{\bf 3}_s}\,
  \tr(\Delta^\dagger \Delta)_{{\bf 3}_a}
 \Bigr)_{\bf 1}
\nonumber\\
&&\hspace*{10mm}
{}
 +
 \lambda_{2\delta\Delta 1}\,
 \tr( \delta^\dagger \delta )\,
 \tr( \Delta^\dagger \Delta )_{\bf 1}
 +
 \lambda_{2\delta\Delta 2}\,
 ( \delta_{\beta\alpha}^\ast \delta_{\omega\gamma} )\,
 ( \Delta_{\beta\alpha} \Delta_{\omega\gamma}^\ast )_{\bf 1}
\nonumber\\
&&\hspace*{10mm}
{}
 +
 \left\{
  \lambda_{2\delta\Delta 3}^\prime\,
  ( \delta_{\beta\alpha}^\ast \delta_{\omega\gamma}^\ast )\,
  \bigl[ \Delta_{\beta\alpha} \Delta_{\omega\gamma} \bigr]_{\bf 1}
  +
  \text{h.c.}
 \right\}
\nonumber\\
&&\hspace*{10mm}
{}
 +
 \left\{
  \lambda_{2\delta\Delta s}^\prime\,
  \delta_{\beta\alpha}^\ast\,
  \bigl[
   \Delta_{\beta\alpha}
   ( \Delta_{\omega\gamma}^\ast \Delta_{\omega\gamma} )_{{\bf 3}_s}
  \bigr]_{\bf 1}
  +
  \text{h.c.}
 \right\}
\nonumber\\
&&\hspace*{10mm}
{}
 +
 \left\{
  \lambda_{2\delta\Delta a}^\prime\,
  \delta_{\beta\alpha}^\ast\,
  \bigl[
   \Delta_{\beta\alpha}
   ( \Delta_{\omega\gamma}^\ast \Delta_{\omega\gamma} )_{{\bf 3}_a}
  \bigr]_{\bf 1}
  +
  \text{h.c.}
 \right\} ,
\end{eqnarray}

\begin{eqnarray}
V_3
&=&
 \frac{1}{\,2\,}\, \lambda_{3\delta}\,
 \left\{
  \bigl[ \tr( \delta^\dagger \delta ) \bigr]^2
  -
  \tr\bigl( \bigl[ \delta^\dagger \delta \bigr]^2 \bigr)
 \right\}
\nonumber\\
&&\hspace*{5mm}
{}+
 \frac{1}{\,2\,}\, \lambda_{3\Delta}\,
 \left\{
  \bigl[ \tr( \Delta^\dagger \Delta )_{\bf 1} \bigr]^2
  -
  \tr\bigl( \bigl[ ( \Delta^\dagger \Delta )_{\bf 1} \bigr]^2 \bigr)
 \right\}
\nonumber\\
&&\hspace*{5mm}
{}+
 \frac{1}{\,2\,}\, \lambda_{3\Delta p}\,
 \Bigl\{
  \tr( \Delta^\dagger \Delta )_{{\bf 1}^\prime}\,
  \tr( \Delta^\dagger \Delta )_{{\bf 1}^\dprime}
  -
  \tr\bigl(
      ( \Delta^\dagger \Delta )_{{\bf 1}^\prime}\,
      ( \Delta^\dagger \Delta )_{{\bf 1}^\dprime}
     \bigr)
 \Bigr\}
\nonumber\\
&&\hspace*{5mm}
{}+
 \frac{1}{\,2\,}\, \lambda_{3\Delta ss}\,
 \Bigl\{
  \bigl(
   \tr( \Delta^\dagger \Delta )_{{\bf 3}_s}
   \tr( \Delta^\dagger \Delta )_{{\bf 3}_s}
  \bigr)_{\bf 1}
  -
  \tr\bigl(
      ( \Delta^\dagger \Delta )_{{\bf 3}_s}
      ( \Delta^\dagger \Delta )_{{\bf 3}_s}
     \bigr)_{\bf 1}
 \Bigr\}
\nonumber\\
&&\hspace*{5mm}
{}+
 \frac{1}{\,2\,}\, \lambda_{3\Delta aa}\,
 \Bigl\{
  \bigl(
   \tr( \Delta^\dagger \Delta )_{{\bf 3}_a}
   \tr( \Delta^\dagger \Delta )_{{\bf 3}_a}
  \bigr)_{\bf 1}
  -
  \tr\bigl(
      ( \Delta^\dagger \Delta )_{{\bf 3}_a}
      ( \Delta^\dagger \Delta )_{{\bf 3}_a}
     \bigr)_{\bf 1}
 \Bigr\}
\nonumber\\
&&\hspace*{5mm}
{}+
 \frac{1}{\,2\,}\, i \lambda_{3\Delta sa}\,
 \Bigl\{
  \bigl(
   \tr( \Delta^\dagger \Delta )_{{\bf 3}_s}\,
   \tr( \Delta^\dagger \Delta )_{{\bf 3}_a}
  \bigr)_{\bf 1}
  -
  \tr\bigl(
      ( \Delta^\dagger \Delta )_{{\bf 3}_s}\,
      ( \Delta^\dagger \Delta )_{{\bf 3}_a}
     \bigr)_{\bf 1}
 \Bigr\}\nonumber\\
&&\hspace*{5mm}
{}+
 \frac{1}{\,2\,}\, \lambda_{3 \delta\Delta 1}\,
 \Bigl\{
  \tr( \delta^\dagger \delta )\,
  \tr( \Delta^\dagger \Delta )_{\bf 1}
  -
  \tr\bigl(
      ( \delta^\dagger \delta )\,
      ( \Delta^\dagger \Delta )_{\bf 1}
     \bigr)
 \Bigr\}\nonumber\\
&&\hspace*{5mm}
{}+
 \frac{1}{\,2\,}\, \lambda_{3 \delta\Delta 2}\,
 \Bigl\{
  \delta^\ast_{\beta\alpha}
  ( \Delta_{\beta\alpha} \Delta^\ast_{\omega\gamma} )_{\bf 1}
  \delta_{\omega\gamma}
  -
  \tr\bigl(
      \delta^\dagger
      ( \Delta \Delta^\dagger )_{\bf 1}
      \delta
     \bigr)
 \Bigr\}\nonumber\\
&&\hspace*{5mm}
{}+
 \Bigl\{
  \frac{1}{\,2\,}\, \lambda_{3 \delta\Delta 3}^\prime\,
   \Bigl(
    ( \delta^\ast_{\beta\alpha} \delta^\ast_{\omega\gamma} )\,
    \bigl[ \Delta_{\beta\alpha} \Delta_{\omega\gamma} \bigr]_{\bf 1}
    -
    \delta^\ast_{\beta\alpha} \delta^\ast_{\omega\gamma}
    \bigl[ \Delta_{\beta\gamma} \Delta_{\omega\alpha} \bigr]_{\bf 1}
   \Bigr)
  + \text{h.c.}
 \Bigr\}\nonumber\\
&&\hspace*{5mm}
{}+
 \Bigl\{
  \frac{1}{\,2\,}\, \lambda_{3 \delta\Delta s}^\prime
  \Bigl(
   \delta^\ast_{\beta\alpha}\,
   \bigl[
    \Delta_{\beta\alpha}
    ( \Delta_{\omega\gamma}^\ast \Delta_{\omega\gamma} )_{{\bf 3}_s}
   \bigr]_{\bf 1}
   -
   \delta^\ast_{\beta\alpha}\,
   \bigl[
    \Delta_{\beta\gamma}
    ( \Delta_{\omega\gamma}^\ast \Delta_{\omega\alpha} )_{{\bf 3}_s}
   \bigr]_{\bf 1}
  \Bigr)
  + \text{h.c.}
 \Bigr\}\nonumber\\
&&\hspace*{5mm}
{}+
 \Bigl\{
  \frac{1}{\,2\,}\, \lambda_{3 \delta\Delta a}^\prime
%  \Bigl(
   \delta^\ast_{\beta\alpha}\,
   \bigl[
    \Delta_{\beta\alpha}
    ( \Delta_{\omega\gamma}^\ast \Delta_{\omega\gamma} )_{{\bf 3}_a}
   \bigr]_{\bf 1}
%   -
%   \delta^\ast_{\beta\alpha}\,
%   \bigl[
%    \Delta_{\beta\gamma}
%    ( \Delta_{\omega\gamma}^\ast \Delta_{\omega\alpha} )_{{\bf 3}_a}
%   \bigr]_{\bf 1}
%  \Bigr)
  + \text{h.c.}
 \Bigr\} ,
\end{eqnarray}

\begin{eqnarray}
&&
V_\mu
=
 \frac{1}{\sqrt{2}}\,
 \mu_\delta\,
 \bigl[ \Phi_\alpha \Phi_\beta \bigr]_{\bf 1}\,
 (i\sigma^2 \delta^\dagger)_{\alpha\beta}
 +
 \frac{1}{\sqrt{2}}\,
 \mu_\Delta\,
 \Bigl(
  ( \Phi_\alpha \Phi_\beta )_{{\bf 3}_s}\,
  (i\sigma^2 \Delta^\dagger)_{\alpha\beta}
 \Bigr)_{\bf 1}
 +
 \text{h.c.} 
\end{eqnarray}
 Note that $V_3$ can be rewritten
in term of the determinant
by using
\begin{eqnarray}
&&
 \tr( \Delta_A^\dagger \Delta_B )\,
 \tr( \Delta_C^\dagger \Delta_D ) 
 -
 \tr\left(
     ( \Delta_A^\dagger \Delta_B )
     ( \Delta_C^\dagger \Delta_D )
    \right)
\nonumber\\
&&\hspace*{10mm}
= \begin{vmatrix}
   ( \Delta_A^\dagger \Delta_B )_{11}
    & ( \Delta_A^\dagger \Delta_B )_{12}\\
   ( \Delta_C^\dagger \Delta_D )_{21}
    & ( \Delta_C^\dagger \Delta_D )_{22}
  \end{vmatrix}
 +
  \begin{vmatrix}
   ( \Delta_C^\dagger \Delta_D )_{11}
    & ( \Delta_C^\dagger \Delta_D )_{12}\\
   ( \Delta_A^\dagger \Delta_B )_{21}
    & ( \Delta_A^\dagger \Delta_B )_{22}
  \end{vmatrix} .
\end{eqnarray}

%%%%%%%% Neutral higgs mass matrix %%%%%%%%
\section{Masses of triplet-like neutral Higgs bosons}
\label{app:neutral}

 Since fields in the $T$-diagonal basis have $Z_3$-charges,
they can not be the mass eigenstates for neutral particles
while they turn out to be the ones for charged particles.
 We show here that the mass eigenstates
of the triplet-like neutral Higgs bosons
just for the completeness,
which seem the most complicated ones in the A4HTM\@.
 We assume that
there is no large mixing between
triplet and doublet fields,
which is possible with small triplet vev's in principle
(See \cite{Akeroyd:2010je} for the case in the HTM).
 The squared mass matrix for
$(\text{Re}(\Delta_x), \cdots, \text{Re}(\delta),
\text{Im}(\Delta_x), \cdots, \text{Im}(\delta))$ is given by
\begin{eqnarray}
M_{T0}^2
&\equiv&
 \begin{pmatrix}
  M_{\text{TCPC}}^2  & M_{\text{TCPV}}^2\\
  (M_{\text{TCPV}}^2)^T & M_{\text{TCPC}}^2
 \end{pmatrix} ,
\end{eqnarray}
\begin{eqnarray}
M_{\text{TCPC}}^2
&\equiv&
 \begin{pmatrix}
  M_{\Delta 45p}^2
   & \frac{1}{\,3\,} v^2 \lambda_{\Delta ss 45p}
   & \frac{1}{\,3\,} v^2 \lambda_{\Delta ss 45p}
   & \frac{1}{\,3\,} v^2 \text{Re}( \lambda_{s 45p}^\prime )\\
  \frac{1}{\,3\,} v^2 \lambda_{\Delta ss 45p}
   & M_{\Delta 45p}^2
   & \frac{1}{\,3\,} v^2 \lambda_{\Delta ss 45p}
   & \frac{1}{\,3\,} v^2 \text{Re}( \lambda_{s 45p}^\prime )\\
  \frac{1}{\,3\,} v^2 \lambda_{\Delta ss 45p}
   & \frac{1}{\,3\,} v^2 \lambda_{\Delta ss 45p}
   & M_{\Delta 45p}^2
   & \frac{1}{\,3\,} v^2 \text{Re}( \lambda_{s 45p}^\prime )\\
  \frac{1}{\,3\,} v^2 \text{Re}( \lambda_{s 45p}^\prime )
   & \frac{1}{\,3\,} v^2 \text{Re}( \lambda_{s 45p}^\prime )
   & \frac{1}{\,3\,} v^2 \text{Re}( \lambda_{s 45p}^\prime )
   & M_{\delta 45p}^2
 \end{pmatrix} ,
\end{eqnarray}
\begin{eqnarray}
M_{\text{TCPV}}^2
\equiv
 \begin{pmatrix}
  0
   & - \frac{1}{\,3\,} v^2 \lambda_{\Delta sa 45p}
   & - \frac{1}{\,3\,} v^2 \lambda_{\Delta sa 45p}
   & \frac{1}{\,3\,} v^2 \text{Im}( \lambda_{s 45p}^\prime )\\
  - \frac{1}{\,3\,} v^2 \lambda_{\Delta sa 45p}
   & 0
   & - \frac{1}{\,3\,} v^2 \lambda_{\Delta sa 45p}
   & \frac{1}{\,3\,} v^2 \text{Im}( \lambda_{s 45p}^\prime )\\
  - \frac{1}{\,3\,} v^2 \lambda_{\Delta sa 45p}
   & - \frac{1}{\,3\,} v^2 \lambda_{\Delta sa 45p}
   & 0
   & \frac{1}{\,3\,} v^2 \text{Im}( \lambda_{s 45p}^\prime )\\
  - \frac{1}{\,3\,} v^2 \text{Im}( \lambda_{s 45p}^\prime )
   & - \frac{1}{\,3\,} v^2 \text{Im}( \lambda_{s 45p}^\prime )
   & - \frac{1}{\,3\,} v^2 \text{Im}( \lambda_{s 45p}^\prime )
   & 0
 \end{pmatrix} ,
\end{eqnarray}
\begin{eqnarray}
M_{\delta 45p}^2
\equiv
 M_\delta^2 + \frac{1}{\,2\,} v^2 \lambda_{\delta 45p} , \quad
M_{\Delta 45p}^2
\equiv
 M_\Delta^2 + \frac{1}{\,2\,} v^2 \lambda_{\Delta 45p} ,
\end{eqnarray}
where $\lambda_{45p}$ are defined as
$\lambda_4 + \lambda_5$ for each subscripts.
 The squared mass matrix $M_{T0}^2$ can be diagonalized
as $O_{T0} M_{T0}^2 O_{T0}^T$
by the orthogonal matrix $O_{T0}$:
\begin{eqnarray}
%O_{\text{neut}}
O_{T0}
\equiv
 O_{Ts} O_{\Delta sa}
 \begin{pmatrix}
  O_{\text{TCPC}} & 0_{4\times 4}\\
    0_{4\times 4} & O_{\text{TCPC}}
 \end{pmatrix} ,
\end{eqnarray}
\begin{eqnarray}
O_{\text{TCPC}}
&\equiv&
 \frac{1}{\sqrt{3}}
 \begin{pmatrix}
  \frac{1}{\sqrt{2}}
   & \frac{1}{\sqrt{2}}
   & 0
   & 0\\
  -\frac{i}{\sqrt{2}}
   & \frac{i}{\sqrt{2}}
   & 0
   & 0\\
  0
   & 0
   & \cos\theta_{T0}
   & \sin\theta_{T0}\\
  0
   & 0
   & -\sin\theta_{T0}
   & \cos\theta_{T0}
 \end{pmatrix}
 \begin{pmatrix}
  1
   & \omega
   & \omega^2
   & 0\\
  1
   & \omega^2
   & \omega
   & 0\\
  1
   & 1
   & 1
   & 0\\
  0
   & 0
   & 0
   & \sqrt{3}
 \end{pmatrix} ,
\end{eqnarray}
\begin{eqnarray}
O_{Ts}
&\equiv&
 \begin{pmatrix}
  1
   & 0
   & 0
   & 0
   & 0
   & 0
   & 0
   & 0\\
  0
   & 1
   & 0
   & 0
   & 0
   & 0
   & 0
   & 0\\
  0
   & 0
   & \cos\theta_{Ts}
   & 0
   & 0
   & 0
   & 0
   & \sin\theta_{Ts}\\
  0
   & 0
   & 0
   & \cos\theta_{Ts}
   & 0
   & 0
   & \sin\theta_{Ts}
   & 0\\
  0
   & 0
   & 0
   & 0
   & 1
   & 0
   & 0
   & 0\\
  0
   & 0
   & 0
   & 0
   & 0
   & 1
   & 0
   & 0\\
  0
   & 0
   & 0
   & -\sin\theta_{Ts}
   & 0
   & 0
   & \cos\theta_{Ts}
   & 0\\
  0
   & 0
   & -\sin\theta_{Ts}
   & 0
   & 0
   & 0
   & 0
   & \cos\theta_{Ts}\\
 \end{pmatrix} ,
\end{eqnarray}
\begin{eqnarray}
O_{\Delta sa}
\equiv
 \begin{pmatrix}
  \frac{1}{\sqrt{2}}
   & 0
   & 0
   & 0
   & 0
   & \frac{1}{\sqrt{2}}
   & 0
   & 0\\
  0
   & \frac{1}{\sqrt{2}}
   & 0
   & 0
   & \frac{1}{\sqrt{2}}
   & 0
   & 0
   & 0\\
  0
   & 0
   & 1
   & 0
   & 0
   & 0
   & 0
   & 0\\
  0
   & 0
   & 0
   & 1
   & 0
   & 0
   & 0
   & 0\\
  0
   & -\frac{1}{\sqrt{2}}
   & 0
   & 0
   & \frac{1}{\sqrt{2}}
   & 0
   & 0
   & 0\\
  -\frac{1}{\sqrt{2}}
   & 0
   & 0
   & 0
   & 0
   & \frac{1}{\sqrt{2}}
   & 0
   & 0\\
  0
   & 0
   & 0
   & 0
   & 0
   & 0
   & 1
   & 0\\
  0
   & 0
   & 0
   & 0
   & 0
   & 0
   & 0
   & 1
 \end{pmatrix} .
\label{eq:ODsa}
\end{eqnarray}
 The mixing angles are defined as
\begin{eqnarray}
\tan{2\theta_{T0}}
&\equiv&
 \frac{ 2\sqrt{3}\, v^2 \text{Re}( \lambda_{s 45p}^\prime ) }
      { 3M_{\Delta 45p}^2 - 3M_{\delta 45p}^2 + 2\lambda_{\Delta ss 45p} v^2 } ,\\
\tan{2\theta_{Ts}}
&\equiv&
 \frac{ 2\sqrt{3}\, v^2 \text{Im}( \lambda_{s 45p}^\prime ) }
      { 
       (
        3M_{\Delta 45p}^2 - 3M_{\delta 45p}^2
        + 2 v^2 \lambda_{\Delta ss 45p}
       )
       \cos{2\theta_{T0}}
       +
       2\sqrt{3}\, v^2 \text{Re}( \lambda_{s 45p}^\prime )
       \sin{2\theta_{T0}}
      } .
\end{eqnarray}
 Note that maximal mixings in $O_{\Delta sa}$ appear only for
the case that the squared triplet vev's (which we ignored here)
are much smaller than $v^2 \lambda_{\Delta sa 45p}$;
 if not, $O_{\Delta sa}$ is almost the unit matrix.

 The mass eigenstates and their masses are obtained as
\begin{eqnarray}
&&
\begin{pmatrix}
 H_{T1}^0 ,
% H_{T2}^0 ,
% H_{T3}^0 ,
 \cdots ,
 H_{T4}^0 ,
 A_{T1}^0 ,
% A_{T2}^0 ,
% A_{T3}^0 ,
 \cdots ,
 A_{T4}^0
\end{pmatrix}^T
\nonumber\\
&&\hspace*{20mm}
=
 O_{T0}
 \begin{pmatrix}
  \text{Re}(\Delta_x^0) ,
%  \text{Re}(\Delta_y^0) ,
%  \text{Re}(\Delta_z^0) ,
  \cdots ,
  \text{Re}(\delta^0) ,
  \text{Im}(\Delta_x^0) ,
%  \text{Im}(\Delta_y^0) ,
%  \text{Im}(\Delta_z^0) ,
  \cdots ,
  \text{Im}(\delta^0)
 \end{pmatrix}^T ,
\end{eqnarray}
\begin{eqnarray}
m_{H_{T1}^0}^2
=
m_{A_{T1}^0}^2
&=&
 M_{\Delta 45p}^2
 - \frac{1}{\,3\,} v^2
   \left(
    \lambda_{\Delta ss 45p}
    + \sqrt{3} \lambda_{\Delta sa 45p}
   \right) ,
\label{eq:m10}\\
m_{H_{T2}^0}^2
=
m_{A_{T2}^0}^2
&=&
 M_{\Delta 45p}^2
 - \frac{1}{\,3\,} v^2
   \left(
    \lambda_{\Delta ss 45p}
    - \sqrt{3} \lambda_{\Delta sa 45p}
   \right) ,
\label{eq:m20}\\
m_{H_{T3}^0}^2
=
m_{A_{T3}^0}^2
&=&
 \frac{1}{\,6\,}
 \left(
  M_{\delta 45p}^2
  + M_{\Delta 45p}^2
  + 2 \lambda_{\Delta ss 45p} v^2
  - 3 \Delta m^2_0
 \right) ,
\label{eq:m30}\\
m_{H_{T3}^0}^2
=
m_{A_{T3}^0}^2
&=&
 \frac{1}{\,6\,}
 \left(
  M_{\delta 45p}^2
  + M_{\Delta 45p}^2
  + 2 \lambda_{\Delta ss 45p} v^2
  + 3 \Delta m^2_0
 \right) ,
\label{m40}\\
3\Delta m^2_0
&\equiv&
 \left\{
  \left(
   3M_{\Delta 45p}
   - 3M_{\delta 45p}
   + 2\lambda_{\Delta ss 45p} v^2
  \right)^2
  +
  12 | \lambda_{s 45p}^\prime |^2 v^4
 \right\}^{\frac{1}{\,2\,}} .
\label{eq:dm0}
\end{eqnarray}
 Of course,
$H_{Ti}^0$ and $A_{Ti}^0$ become
the CP-even and odd neutral Higgs bosons, respectively,
if $M_{\text{TCPV}}^2$ vanishes.
 It is clear that eq.~(\ref{eq:m10})-(\ref{eq:dm0})
can be given by replacing
$\lambda_5$ with $-\lambda_5$
in eq.~(\ref{eq:m1pp})-(\ref{eq:dmpp}).

\end{document}